\documentclass[aps,prd,psfig,amsmath,amsfonts,groupedaddress,eqsecnum,floatfix]{revtex4}

\usepackage{graphicx,epsfig}
\usepackage{bm}

\newcommand{\be}{\begin{equation}}
\newcommand{\ee}{\end{equation}}
\newcommand{\bea}{\begin{eqnarray}}
\newcommand{\eea}{\end{eqnarray}}
\newcommand{\ba}{\begin{array}}
\newcommand{\ea}{\end{array}}
\newcommand{\beqn}{\begin{eqnarray*}}
\newcommand{\eeqn}{\end{eqnarray*}}

\newcommand{\f}[2]{\frac{#1}{#2}}

\begin{document}

\nopagebreak

\title{A post-Newtonian diagnosis of quasiequilibrium configurations of\\
  neutron star-neutron star and neutron star-black hole binaries}

\author{Emanuele Berti
}
\email[]{berti@wugrav.wustl.edu}

\author{Sai Iyer}
\email[]{sai@wugrav.wustl.edu}

\author{Clifford M. Will}
\email[]{cmw@wuphys.wustl.edu}

\affiliation{\it McDonnell Center for the Space Sciences, 
Department of Physics, Washington University, 
St. Louis, Missouri 63130, USA}

\begin{abstract}
\noindent

We use a post-Newtonian diagnostic tool to examine numerically
generated quasiequilibrium initial data
sets for non-spinning double neutron star and neutron star-black hole
binary systems.  The PN equations include the effects of tidal
interactions, parametrized by the compactness of the neutron stars and
by suitable values of ``apsidal'' constants, which measure the degree
of distortion of stars subjected to tidal forces.
We find that the post-Newtonian diagnostic 
agrees well with the double neutron star initial data, typically to better
than half a percent except where tidal distortions are becoming
extreme.  We show that the differences
could be interpreted as
representing small residual eccentricity in the initial orbits.  In
comparing the diagnostic with preliminary 
numerical data on neutron star-black hole binaries, we find less
agreement.

\end{abstract}
\maketitle

\section{Introduction and Summary}
\label{sec:intro}

Advances in numerical general relativity have made it possible to
solve Einstein's equations for the astrophysically important problem
of the inspiral and merger of binary systems of compact bodies (black
holes or neutron stars) with unprecedented accuracy and robustness,
both in the initial data obtained from Einstein's equations, and in
the subsequent time evolution of the system.  Work is ongoing to
stitch together results from post-Newtonian theory, which accurately
describes the early part of the inspiral, with numerical relativity,
which describes the final few orbits, merger and ringdown of the final
black hole, in an effort to develop a complete description of the
evolution and gravitational-wave emission from such systems that can
be used effectively in analysis of data from gravitational-wave
interferometers~\cite{bcp07,bertijena,baker07,hannam07}.

As part of this program to link post-Newtonian theory
with numerical relativity,
we proposed and developed a diagnostic tool, based on the
post-Newtonian (PN) approximation, for analysing numerical relativity
initial data sets  for compact binaries
\cite{MW1,MW2,Berti:2006bj}.  Recall that the post-Newtonian
approximation
is effectively an expansion of Einstein's theory in powers of
a small parameter
$\epsilon \sim (v/c)^2 \sim Gm/rc^2$. 

Numerical data sets, obtained from solutions of the
initial-value equations of general relativity, are generally
configured to represent the
quasicircular orbit of a compact binary system late in its
inspiral phase, when the bodies have only a few orbits remaining before
merger.  Such a quasicircular orbit 
(circular, apart from the
radiation-induced inspiral)
is the expected end-point of evolution
of a compact binary system under the circularizing effect of gravitational
radiation reaction in the absence of external perturbations.  

The PN diagnostic provides analytic
expressions for the total binding energy 
(the difference between the total gravitational mass of the system and
that of the two stars in isolation) 
and angular
momentum of the system as a function of its orbital angular velocity, and
allows for arbitrary mass ratios, spins, finite-size effects
such as tidal interactions, and a non-zero orbital eccentricity.
We found that the PN expressions for circular orbits 
agreed with the numerical relativity results
surprisingly well (see also~\cite{luc}), 
even for orbital separations in a highly relativistic
regime where $\epsilon$ is not so small, but 
that there were systematic differences that could not be
blithely attributed to errors in either the PN approximation or the
numerical data.  

For binary black hole initial data, for example, we found 
\cite{MW1,Berti:2006bj} that our diagnostic gave a better fit to the numerical
data from several groups if the orbital eccentricity were allowed
to be non-zero,
with values as large as 0.03.  Here the bodies were viewed as
residing initially at the
apocenter of the orbit, in keeping with the constraint built
into the numerical initial data that the bodies be moving
perpendicularly to their separation.   
And indeed, subsequent calculations of the time evolution of these
initial data using numerical relativity showed that the orbits {\em were}
slightly eccentric, with values of eccentricity agreeing in order of
magnitude
with those predicted by the PN diagnostic 
(see, eg. \cite{baker07b}).  Efforts are now being made to ``tweak'' the initial
data for binary black holes 
in order to achieve orbits that more nearly approximate
the expected quasicircular orbits 
~\cite{pfeiffer07,husa07,grigsby}.

In this paper we turn our attention to binary systems containing non-rotating
neutron stars (NS).  In \cite{MW2} we applied the diagnostic to one neutron
star binary initial data set for corotating stars obtained by Miller {\em et
  al.} \cite{miller}, but here we focus on an extensive set of {\em
  non-spinning} binary neutron star initial data obtained by Taniguchi {\em et
  al.}~\cite{L1,L2,L3}, including both equal-mass and unequal-mass systems;
and on preliminary initial data sets for neutron star-black hole binaries
obtained by Taniguchi {\em et al.}
\cite{Taniguchi:2006yt,Taniguchi:2007xm,Taniguchi:PC} and Grandcl\'ement
\cite{Grandclement:2006ht}.

While for black holes, tidal effects are negligible at the separations
in question, for neutron stars, 
tidally induced
distortions must be taken into account (see \cite{MW2} for
discussion). 
These effects depend on the size of the
neutron star, as encoded in a ``compactness'' factor, $M/R$, where $M$ and
$R$ represent mass and radius, and on a set of ``apsidal constants'', $k_l$,
which measure the degree of distortion of the body as a result of an
external tidal force, for each multipole order $l$.  These are also known as
``Love numbers'' in other contexts.  The apsidal constants depend on the
equation of state (EOS) used to model the neutron star interiors.

For given values of the masses,
compactness parameters, apsidal constants and eccentricity, the
PN
expressions for binding energy and angular
momentum can be evaluated and compared with the data.
Alternatively, for example, one can leave the eccentricity as a free
parameter and solve for that value that gives the best agreement
between the PN and the numerical results for the given angular
velocity.

\begin{figure*}[htb]
\begin{center}
\begin{tabular}{cc}
\epsfig{file=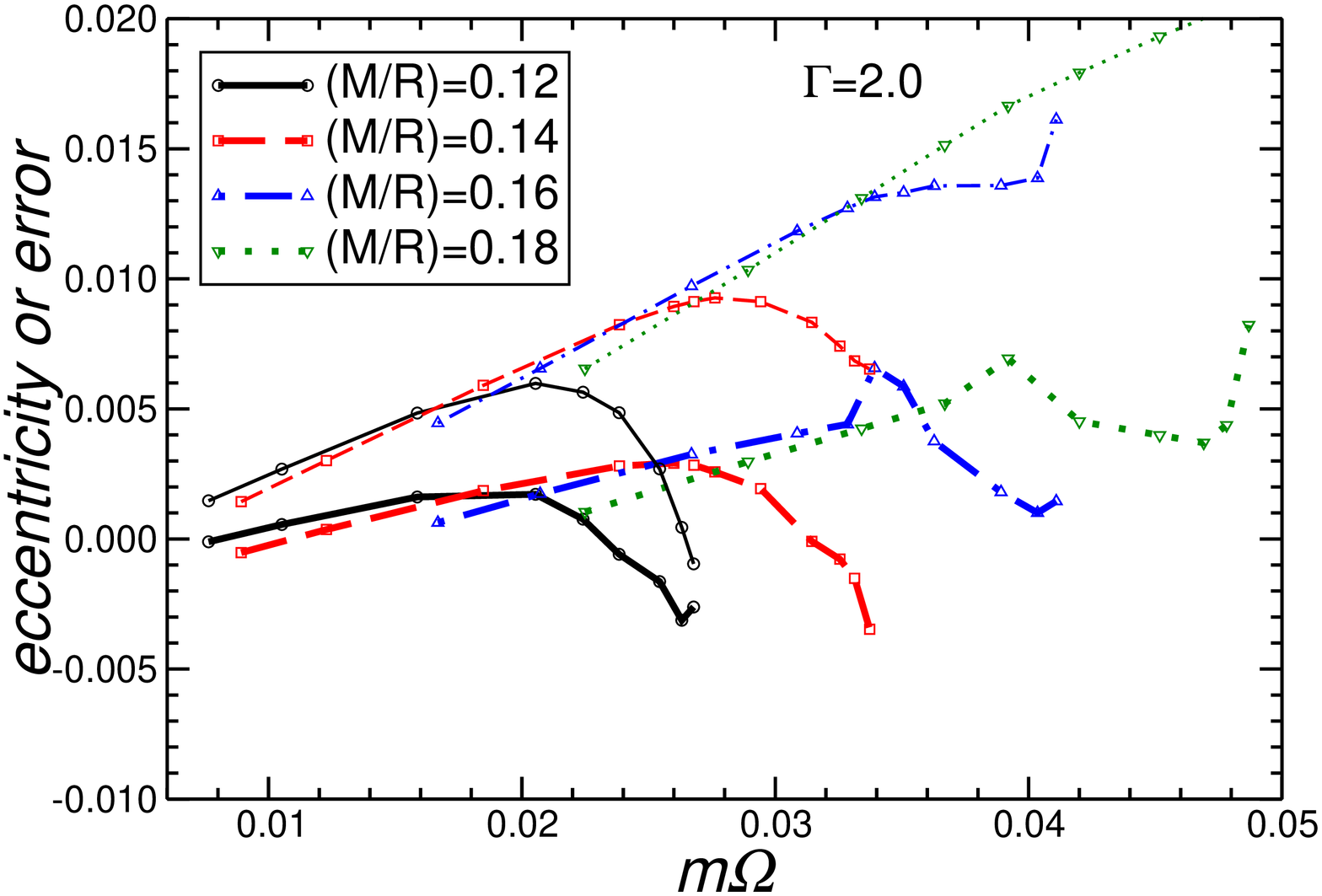,width=6cm,angle=-90} &
\epsfig{file=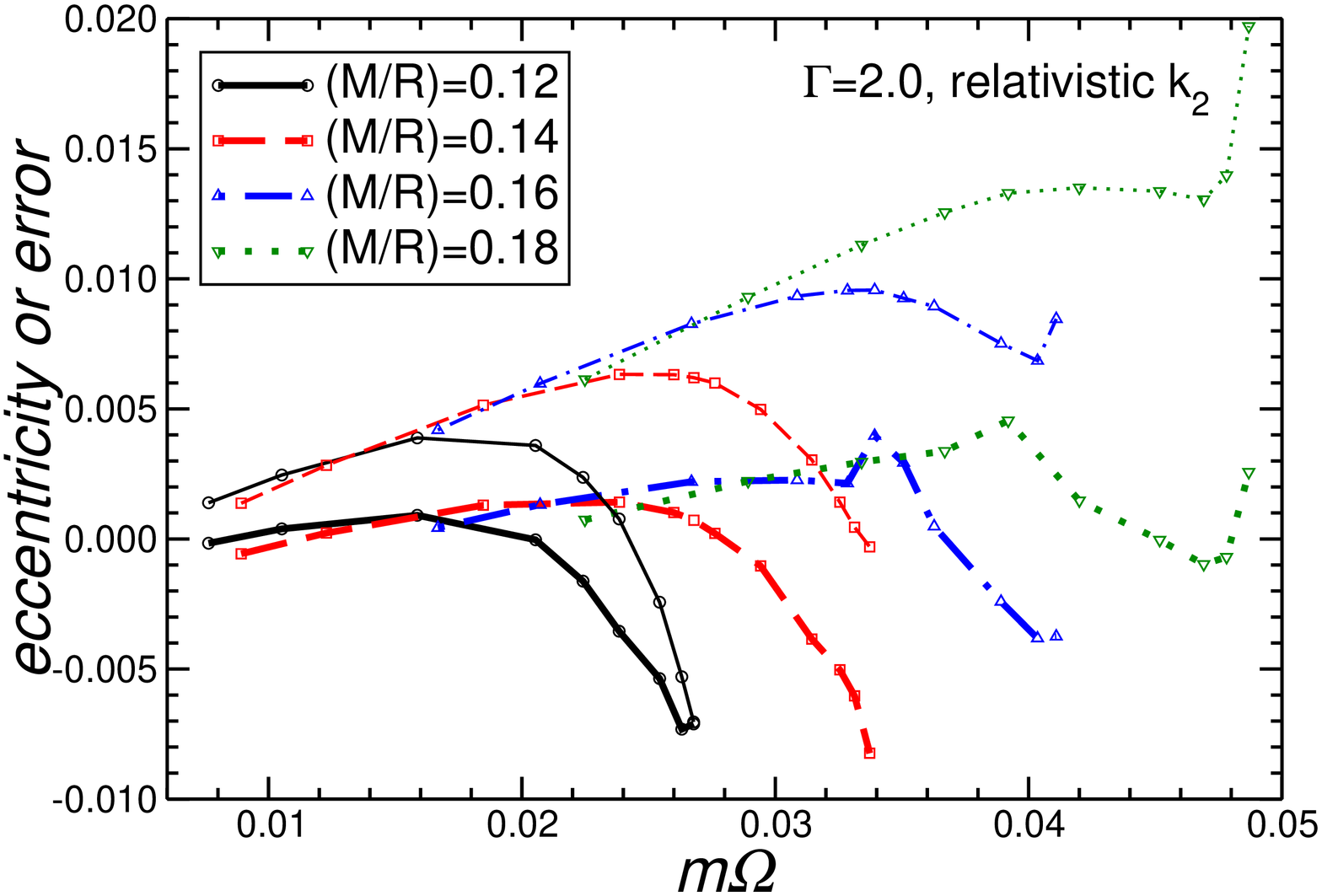,width=6cm,angle=-90} \\
\epsfig{file=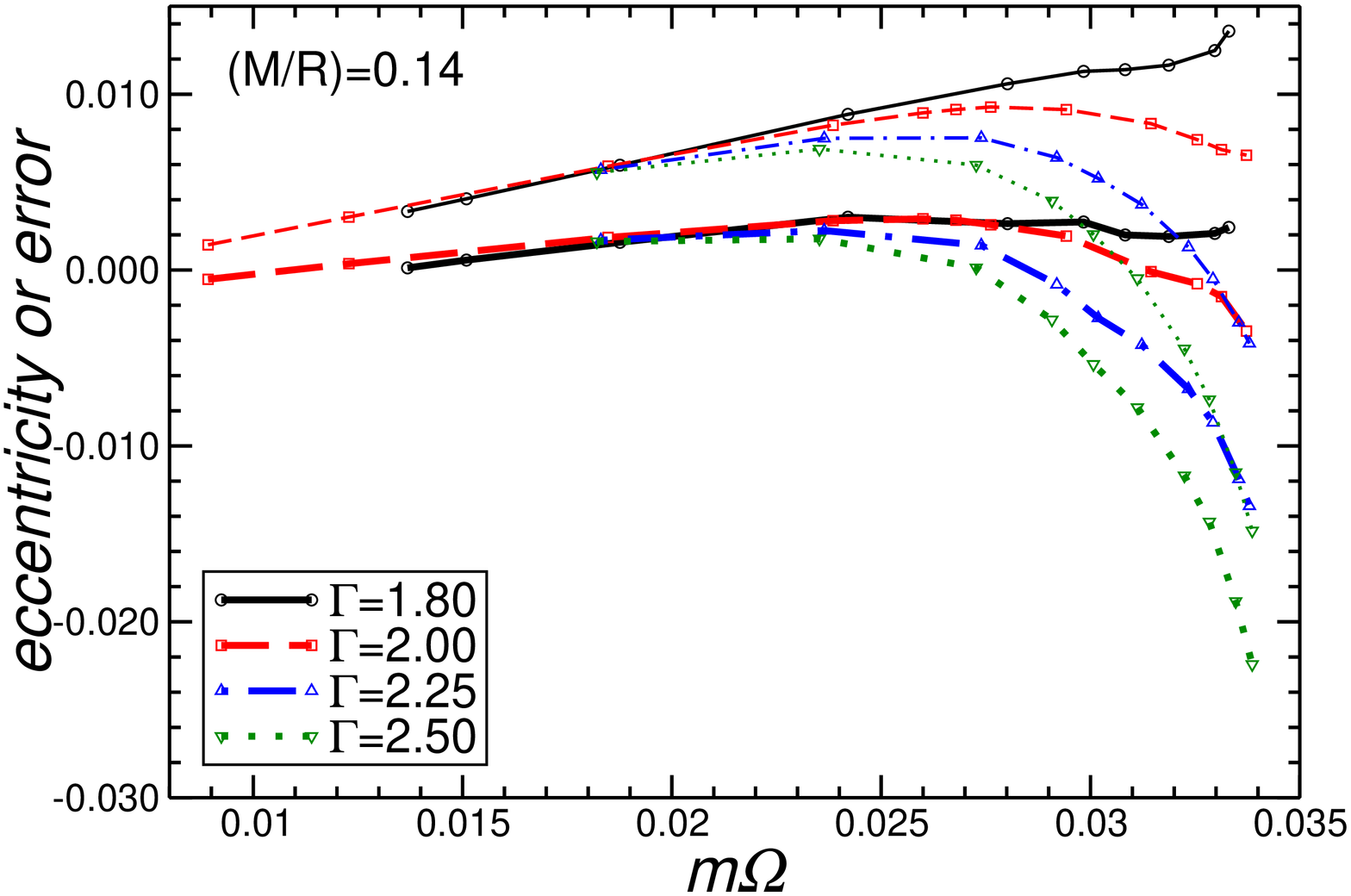,width=6cm,angle=-90} &
\epsfig{file=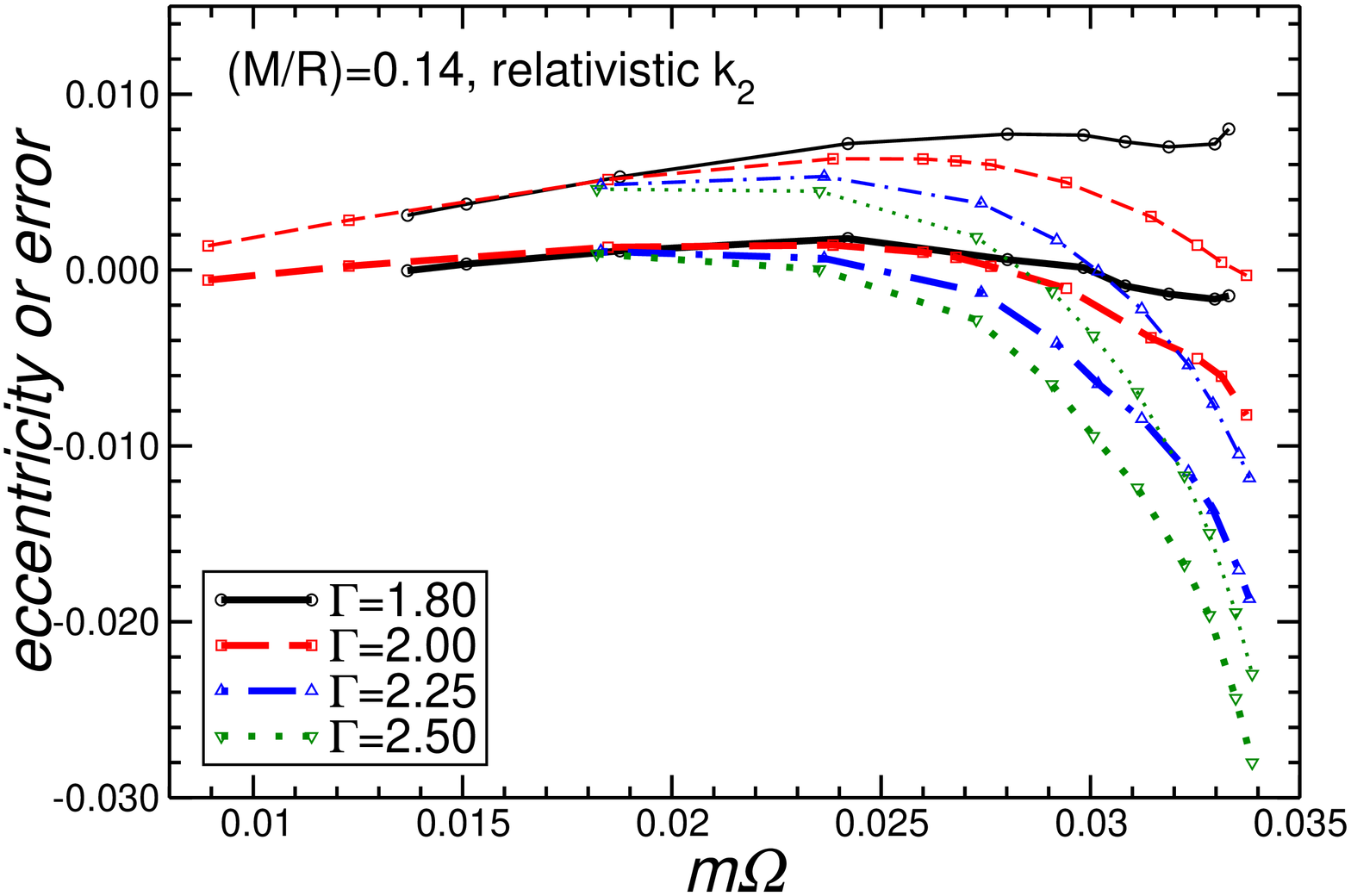,width=6cm,angle=-90} \\
\end{tabular}
\caption{Best-fit eccentricity estimated from the energy (thick lines) and the
  angular momentum (thin lines) for irrotational, equal-mass neutron
  star-neutron star binaries. Top: we fix $\Gamma=2.00$ and consider
  different values of the compactness. Bottom: we fix $M/R=0.14$ and
  consider different values of $\Gamma$. Results on the left use the Newtonian
  apsidal constants, results on the right use the ``relativistically
  corrected'' apsidal constants of
  Eq.~(\ref{k2rel}). \label{fig:eccentricity}}
\end{center}
\end{figure*}

In comparing with the non-spinning NS-NS initial data sets 
of~\cite{L1,L2,L3},  
our main results can be summarized using Fig. \ref{fig:eccentricity}.
These figures show the eccentricity required to match PN theory with
the numerical data for equal-mass systems, 
for both binding energy (thick lines) and angular
momentum (thin lines), for various equations of state, as represented
by the adiabatic index $\Gamma$, and for various values of the neutron
star compactness $M/R$.  The panels on the left use values for the
apsidal constants $k_2$ and $k_3$
calculated using Newtonian theory, while those on
the right use values for the dominant $k_2$ apsidal constant estimated 
from rotating relativistic stellar models (see Appendix
\ref{sec:rotcorr}).   The figures are plotted against $m\Omega$, where
$m$ is the total gravitational mass of the two isolated stars and
$\Omega$ is the orbital angular velocity.  The left-hand end
of each plot
corresponds to the ``Newtonian'' limit, where the stars are farther
apart; on the right-hand end, the sequences terminate when the stars
are within 25 to 40 percent of touching, where cusps in the surface
shape are beginning to develop, signalling the onset of mass
shedding.

As discussed below (Sec.~\ref{sec:pnd-eqs}), 
the inferred eccentricity is also an approximate measure
of the fractional difference between the data and PN theory assuming
circular orbits, so the figures can equally well be interpreted as a measure
of the accuracy of the PN-numerical comparison.

From these figures, together with considerations in following
sections, we summarize our main results.

{\em PN theory gives an excellent fit, better than half a percent in
most cases, to the binding energy 
of the  numerically generated initial data sequences for all but the
closest separations.} 
Tidal effects matter, and
one must use the correct compactness factor and apsidal constants
corresponding to the isolated stars and equations of state
used in the numerical solutions.  

{\em If circular orbits are assumed, there are systematic differences 
between the numerical data and the
PN diagnostic.}  These could be interpreted as being
the result of residual eccentricity in the simulations, albeit at levels
smaller than those inferred from the BH-BH models in~\cite{MW1,Berti:2006bj}.
The differences are larger than can be accounted for either by truncation
errors
in the PN series or by numerical errors in the initial data sets.

{\em When best-fit eccentricities are calculated, the values
inferred from the energy diagnostic are generally smaller
than those inferred from the angular momentum diagnostic.}  The same
phenomenon was seen in the BH-BH diagnostics~\cite{MW1,Berti:2006bj}.  
One possibility is that there are
poorly understood
systematic effects in calculating total angular momentum from numerical initial
data sets, or in fixing the intrinsic angular momenta of the individual stars to
the desired values (zero in this case).

{\em A comparison with preliminary initial data for neutron star-black hole
binaries shows that the required eccentricities are larger than for
NS-NS binaries.}  Furthermore, they increase with the mass ratio 
$m_{\rm BH}/m_{\rm NS}$.  However, because these are early days for
NS-BH simulations, the numerical errors in the models are larger than
in the more mature NS-NS simulations, so these conclusions should be
regarded as tentative.  Our PN diagnostic could be a useful tool in
helping to refine and improve these initial data sets.

Because tidal effects
fall off rapidly with distance and scale with the size of the bodies,
the Newtonian tidal contributions are already 
small, and thus we have argued that post-Newtonian tidal
effects are not needed.  However, because the neutron stars are highly
relativistic, one might question the use of apsidal constants derived
from Newtonian theory.
We have addressed this by constructing models of
isolated, slowly rotating, fully relativistic neutron stars, and by
using the rotationally induced deformation to estimate the
``relativistic'' apsidal constant $k_2$.  The resulting values can
vary by a factor of two over the range of compactness factors
considered, leading to small but noticeable effects in the inferred
eccentricities (right-hand panels of Fig. \ref{fig:eccentricity}).

The remainder of this paper provides details.  In Sec.~\ref{sec:pnd-eqs} we
summarize the post-Newtonian diagnostic equations derived in \cite{MW2},
applicable to both NS-NS and NS-BH non-spinning binaries.  In
Sec.~\ref{sec:QENS} we consider the catalogue of quasiequilibrium NS-NS
sequences computed by the Meudon group \cite{L1,L2,L3}.
Section~\ref{sec:QENSBH} considers NS-BH binaries obtained by Taniguchi {\em
  et al.} \cite{Taniguchi:2006yt,Taniguchi:2007xm,Taniguchi:PC} and
Grandcl\'ement \cite{Grandclement:2006ht}. Section~\ref{sec:Conclusion}
presents concluding remarks.  Appendix \ref{sec:rotcorr} provides details of
the calculation of relativistic corrections to the energy, angular momentum
and quadrupole moment of rotating isolated neutron stars described by a
polytropic equation of state. These calculations were performed using the
Hartle-Thorne slow-rotation code described in \cite{BWMB}, and are used to
estimate ``relativistically corrected'' values of the apsidal constant $k_2$.
Hereafter, we use units in which $G=c=1$.

\section{Equations of the PN diagnostic}
\label{sec:pnd-eqs}

We consider a binary system of bodies of mass $m_1$ and $m_2$, where these
masses denote the total, or gravitational mass of an isolated, non-rotating
body of 
given baryon number and equation of state.  For black holes, the masses denote
irreducible mass.  We also define the total mass, $m \equiv m_1 + m_2$, and 
the reduced mass, $\mu \equiv m_1m_2/m$, along with
$\eta \equiv \mu/m=m_1m_2/m^2$ ($0 < \eta \le 1/4$).
We define two quantities $e$
and $\zeta$, related to the eccentricity and semi-latus rectum of the orbit,
according to:
\begin{eqnarray}
e &\equiv& \frac{ \sqrt{\Omega_p} - \sqrt{\Omega_a} } 
	{ \sqrt{\Omega_p} + \sqrt{\Omega_a} } \,,
\nonumber \\
\zeta \equiv
\frac{m}{p} &\equiv& \left ( \frac{\sqrt{m\Omega_p} +
\sqrt{m\Omega_a}}{2} \right )^{4/3} \,,
	\label{ezeta}
\end{eqnarray}
where $\Omega_p$ is the value of the orbital frequency $\Omega$ where it
passes through a local maximum (pericenter), and $\Omega_a$ is the value of
$\Omega$ where it passes through the {\it next} local minimum (apocenter).
These definitions are discussed and justified in \cite{MW1,MW2}; in
the Newtonian limit, they agree exactly with the standard definitions.
It follows from Eqs. (\ref{ezeta}) that $\zeta$ also has the property
that 
\begin{equation}
\zeta = \left ( \frac{m\Omega_p}{(1+e)^2} \right )^{2/3}
        = \left ( \frac{m\Omega_a}{(1-e)^2} \right )^{2/3} \,.
        \label{zetaproperty}
\end{equation}

The diagnostic was restricted to systems of possibly rotating bodies 
with spin axes
aligned perpendicular to the orbital plane 
(see \cite{MW2}, Eqs. (2.35), (3.6), (3.7), (3.15) and
(3.17) for the complete set of ingredients).  
However, because we will be considering numerical initial data sets
for systems with nonrotating bodies, the formulae simplify
significantly, because there will be no contributions from rotational
kinetic energy or angular momentum, spin-orbit or spin-spin coupling,
effects due to rotational distortions or tidal-rotational coupling,
and so on. 
As a result,
the total binding energy $E$ and total angular momentum $J$ 
of the system can 
be written
\begin{eqnarray}
E &=& E_{\rm Harm} + E_{\rm Tidal} 
\,,
\nonumber \\
J &=& J_{\rm Harm} + J_{\rm Tidal} 
\,.
\label{EJtot}
\end{eqnarray}
where the ``point-mass'' contributions to $E$ and $J$, accurate
to third post-Newtonian (3PN) order, and expressed in
harmonic coordinates, are given by (recall that $\zeta \sim
O(\epsilon)$)
\allowdisplaybreaks{
\begin{subequations}
\begin{eqnarray}
{E}_{\textrm{Harm}}&=&-\frac{1}{2}m \eta(1-e^{2}) \zeta
\left \{ 1 -\left [ \frac{3}{4}+\frac{1}{12}\eta-\left (\frac{1}{12}-\frac{1}{4}\eta\right )e^{2}
\right ]\zeta
\right .
\nonumber\\
 &&
 \left .
 -\left [\frac{27}{8}-\frac{19}{8}\eta+\frac{1}{24}\eta^{2}
-\left (\frac{17}{12}+4\eta+\frac{1}{4}\eta^{2}\right )e^{2}
+\left (\frac{1}{24}+\frac{29}{24}\eta-\frac{1}{8}\eta^{2}\right )e^{4}
\right ]\zeta^{2}
\right.
\nonumber\\
&&
\left.
-\left [ \frac{675}{64}-\left (\frac{34445}{576}-\frac{205}{96}\pi^{2} 
\right)\eta
+\frac{155}{96}\eta^{2}+\frac{35}{5184}\eta^{3}
\right.
\right.
\nonumber\\
&&
\left.
\left.
+\left (\frac{7}{64}-\left (\frac{2369}{576}+\frac{41}{96}\pi^{2}\right )\eta
+\frac{11951}{864}\eta^{2}-\frac{25}{576}\eta^{3}\right )e^{2}
\right.
\right.
\nonumber\\
&&
\left.
\left.
-\left (\frac{815}{576}-\frac{7619}{1728} \eta
-\frac{1499}{288}\eta^{2}-\frac{25}{64}\eta^{3}\right )e^{4}
\right.
\right.
\nonumber\\
&&
\left.
\left.
-\left(\frac{35}{5184}-\frac{143}{192}\eta+\frac{57}{32}\eta^{2}
-\frac{5}{64}\eta^{3}\right )e^{6}
\right ]\zeta^{3} \right \} \,,
\label{Eharm}
\\
{J}_{\textrm{Harm}}&=&\frac{m^2 \eta}{\sqrt{\zeta}}\left 
\{1+\left[ \frac{3}{2}+\frac{1}{6}\eta-\left (
\frac{1}{6}-\frac{1}{2}\eta\right )e^{2}\right ]\zeta
\right.
\nonumber\\
&&
+\left [\frac{27}{8}-\frac{19}{8}\eta+\frac{1}{24}\eta^{2}
+\left (\frac{23}{12}-\frac{31}{6}\eta-\frac{1}{4}\eta^{2}\right )e^{2}
+\left (\frac{1}{24}-\frac{35}{24}\eta
-\frac{1}{8}\eta^2\right )e^{4}\right ]\zeta^{2}
\nonumber\\
&&
+\left [\frac{135}{16}-\left (\frac{6889}{144}-\frac{41}{24}\pi^{2}
\right )\eta
+\frac{31}{24}\eta^{2}
+\frac{7}{1296}\eta^{3}
\right.
\nonumber\\
&& 
+\left (\frac{299}{16}-\left (\frac{10003}{144}-\frac{41}{24}\pi^{2}
\right)\eta
+\frac{3013}{216}\eta^{2}-\frac{5}{144}\eta^{3}\right )e^{2}
\nonumber\\
&& 
\left.\left. 
+ \left (\frac{77}{144}-\frac{6497}{432}\eta+\frac{853}{72}\eta^{2}
+\frac{5}{16}\eta^{3}\right)e^{4}
\right.
\right.
\nonumber\\
&&
\left.
\left.
-\left (\frac{7}{1296}+\frac{1}{16}\eta+\frac{1}{8}\eta^{2}
-\frac{1}{16}\eta^{3}\right )e^{6}
\right ]\zeta^{3}\right \} \,.
\label{Jharm}
\end{eqnarray}
\label{EJharm}
\end{subequations}
}
Expressions for $E$ and $J$ are also available in Arnowitt-Deser-Misner (ADM) 
coordinates \cite{MW2}; however the differences are negligible for the
systems in question. 
The tidal contributions to 
$E$ and $J$ are given by, 
\begin{subequations}
\begin{eqnarray}
{E}_{\rm Tidal} &=&
m \eta (1-e^2) \left [
\frac{1}{18}( 9+ 10 e^2 - 3e^4) B \zeta^6
+ \frac{1}{24} (13 + 49e^2 + 7e^4 -5e^6 ) C \zeta^8
\right ]
\,,
\label{ETRorbit}
\\
{J}_{\rm Tidal} &=&
m^2 \eta \left [ 
 \frac{2}{9} (3+ 10e^2 + 3e^4 ) B \zeta^{9/2}
+ \frac{2}{3} (1 +7e^2 + 7e^4 + e^6 ) C \zeta^{13/2} \right ]
\,,
\label{JTRorbit}
\end{eqnarray}
\label{EJTRorbit}
\end{subequations}
where
\begin{eqnarray}
B &=&  6 \eta \left [ q_1^5 k_2^{(1)} \left (\frac{m_1}{m} \right)^3 
+ q_2^5 k_2^{(2)} \left (\frac{m_2}{m} \right)^3 \right ]\,,
\nonumber \\
C  &=& 8 \eta \left [ q_1^7 k_3^{(1)} \left (\frac{m_1}{m} \right)^5
 +  q_2^7 k_3^{(2)} \left (\frac{m_2}{m} \right)^5 \right ]\,.
\label{BCcoeffs}
\end{eqnarray}
For each body, $q_a \equiv R_a/m_a$ 
is the inverse of the ``compactness parameter'', where
$R_a$ is its radius in harmonic coordinates, 
and $k_2^{(a)}$ and $k_3^{(a)}$ denote
the ``apsidal constants'' for angular harmonics $l=2$ and $l=3$, respectively
(higher harmonics make negligible contributions).
In Newtonian gravity, the apsidal constants
depend only on the harmonic index
$l$ and on the equation of state; their values are shown in 
Appendix \ref{sec:rotcorr}, Table
\ref{tab:apsidaln}.  Notice that the leading contribution to
the tidal terms is of order $\zeta^5$ relative to the Newtonian
contributions to $E$ or $J$, or effectively 5PN order.  This is why a
Newtonian treatment of tides is justified.
We also note that, in calculating $E/m$ and
$J/m^2$, 
only mass ratios are relevant, since, apart from the dependence on
$\zeta$, the expressions depend only
on $\eta$, $m_1/m$ or $m_2/m$.

In comparing this analytical diagnostic with numerical initial data,
one must carefully translate the meaning of the variables in the two
approaches.  The PN masses
$m_a$ are the physically measured masses of the corresponding
isolated body; these correspond directly to the 
ADM masses of the isolated bodies defined in numerical relativity.  
Similarly, numerical
stellar radii are
generally computed in ADM
coordinates, which, for an isolated body corresponds to the isotropic
radial coordinate of the Schwarzschild geometry; we denote this radius
by $R_{ADM}$.  
Alternatively, one could
compute, from the numerical data, the proper 
circumferential or areal radius of the
isolated
star [${\cal C}/2\pi$ or $({\cal A}/4\pi)^{1/2}$], 
which would correspond to the Schwarzschild radial coordinate $R_S$.  The
relationship between the PN harmonic radial coordinate $R_H$ and 
the others is
\begin{equation}
R_H =  R_S - M_{ADM} = R_{ADM} \left ( 1 +
\frac{M_{ADM}^2}{4R_{ADM}^2} \right ) \,.
\label{radialcoord}
\end{equation}
Since for neutron stars, $M_{ADM}/R_{ADM} < 0.2$, the difference
between $R_H$ and $R_{ADM}$ is generally less than one percent, and
is thus negligible for our purposes.
For non-rotating black holes, the relevant radius is $R_H = m_{\rm BH}$, but
since tidal effects are negligible for the separations to be
considered, this will not play a role.

Finally, we note that, inserting Eq. (\ref{zetaproperty})
into the Newtonian part of Eqs. (\ref{EJharm}), we obtain
to leading order,
\bea
E/m &\approx& -\frac{1}{2} \eta (m\Omega)^{2/3} \left[ 1 + \frac{4}{3}e +
O(e^2) \right] + O[(m\Omega)^{4/3}]\,,
\nonumber \\
J/m^2 &\approx& \eta (m\Omega)^{-1/3} \left[ 1 - \frac{2}{3}e +
O(e^2) \right]+ O[(m\Omega)^{1/3}] \,.
\label{EJerror}
\eea
Thus, apart from factors of $4/3$ or $2/3$, the eccentricity
$e$ can be viewed as a measure of the
fractional deviation of the energy and angular momentum from its circular orbit
value for a given $m\Omega$.  If the inferred value of $e$ is
positive,  the binary sytems is at apocenter; if the inferred value is
negative, it really corresponds to a positive eccentricity, but with
the binary system at pericenter. 

\section{Quasiequilibrium binary neutron star initial configurations}
\label{sec:QENS}

\subsection{Numerical Calculations}

The first numerical calculations of initial data for binary neutron stars in
full general relativity were
carried out, for the case of momentarily corotating bodies, by
Baumgarte {\em et al.}
\cite{baumgarte1,baumgarte2} and Marronetti {\em et al.}~\cite{mmw-cor}.  
Miller {\it et al.}
\cite{miller} performed convergence tests and included error bars in
their simulations of corotating neutron stars. 

The first results for the more physically realistic case of 
non-spinning, or irrotational binary 
neutron stars were presented by the Meudon group 
using spectral methods~\cite{L0,L1},
by Marronetti {\em et al.}
using a single-domain finite difference method
in Cartesian coordinates~\cite{mmw-irr}, and by  
Uryu {\em et al.} using quasispherical coordinates~\cite{J1,J2,J3}.
Generalization to other values of the stellar rotation rates was
carried out in~\cite{MS}.  

We will focus on simulations of non-spinning binary
neutron stars by the 
Meudon group in \cite{L1,L2,L3}.  
They assume that the system is in a {\em quasiequilibrium} state, meaning
that, in a reference frame that momentarily rotates with the orbital
angular velocity, the system is stationary (technically implying the
existence of a ``helicoidal Killing vector'' $\partial/\partial t +
\Omega \partial/\partial \phi$).
They assume that the fluid flow is 
irrotational with respect to the global frame 
and that the induced spatial metric is conformally
flat.  
They use a multidomain spectral method with two patches of surface-fitting
coordinates, to solve for the five metric
functions (lapse, conformal factor and shift 3-vector) and the velocity
potential of the fluid. 
The input information is an equation of state
of each star; an initial coordinate
separation $d$; and the central enthalpy (or density) of each star, or alternatively
the baryonic mass of each star.

The equation of state used is a zero-temperature polytrope with
$p = \kappa\rho^\Gamma$, where where $p$ and $\rho$ are the pressure
and mass
density, and $\Gamma$ is the adiabatic index.  Such equations of state
are also parametrized by a ``polytropic index'' $n$, related to
$\Gamma$ by $\Gamma = 1 + 1/n$.
Most groups use ``polytropic'' units. whereby one defines a length scale
\be
R_{\rm poly}=\kappa^{1/[2(\Gamma-1)]}\,,
\ee
and rescales every quantity by $R_{\rm poly}$ to make it
dimensionless.
With this choice, equilibrium models are characterized only by 
$\Gamma$ and by the (dimensionless) central energy density of
each body (see eg.
Sec.~2.6.1 of \cite{stergioulas} for a detailed explanation of these
units).

In Ref.~\cite{L1} 
calculations were
limited to binaries of equal mass and to a single polytropic EOS with
$\Gamma=2$, $M/R=0.14$ and baryonic
mass $M_B=1.625 M_\odot$.  Results were given in their Table III.
Taniguchi and Gourgoulhon ~\cite{L2}, again with $\Gamma=2$,
extended the
calculations to corotating and
non-spinning binaries of varying compactness, and also with 
unequal masses.  
Finally in Ref.~\cite{L3} calculations were presented for
an extensive sample of values of 
$\Gamma$ and of $M/R$,
for both equal and unequal masses.
Comparison of the endpoints of their sequences of models with those of
Uryu {\em et al.}~\cite{J1,J2,J3} showed good agreement~\cite{L3}.  
Our Tables~\ref{tab:emmodels} and~\ref{tab:ummodels} 
summarize schematically the different quasiequilibrium sequences 
considered by the Meudon group,
pointing to the relevant Tables in the original papers \cite{L2,L3}.

\begin{table}[htb]
\centering
\caption{Filled entries mark 
combinations of $\Gamma=1+1/n$ and $M/R$ considered by
  the Meudon group for equal-mass binaries.  For non-spinning
configurations, bullets mark cases for
  which we plotted $E/m$ and $J/m^2$, and 
circles mark cases for which we plotted the
inferred eccentricities. 
  The third and fourth columns give
  the maximum baryonic mass $M_B^{\rm max}$ and maximum compactness
  $(M/R)^{\rm max}$, respectively, for the given polytropic model. ``Cor''
  means corotating stars, ``NoS'' means non-spinning stars.
The last three columns give the Tables and References where
  the corresponding numerical data are tabulated. }
\vskip 12pt
\begin{tabular}{@{}|cccc|ccccccc|cc|c|@{}}
\hline
\multicolumn{4}{|c}{Parameters} &\multicolumn{7}{|c|}{$M/R$} &\multicolumn{2}{|c|}{Table} &Ref.\\
\hline
$\Gamma$ &$n$ &$M_B^{\rm max}$ &$(M/R)^{\rm max}$
&$0.08$ &$0.10$ &$0.12$ &$0.14$ &$0.16$ &$0.18$ &$0.20$ &Cor &NoS & \\
\hline
1.80 &5/4 &0.217 &0.172 &$\bullet$ &$\times$ &$\times$ &$\bullet
\circ$ & & & &X   &XII  &\cite{L3}\\
2.00 &1   &0.180 &0.214 & & &$\bullet \circ$ &\hfil$\circ$ &$\circ$ &$\bullet \circ$ & &I   &III  &\cite{L2}\\
2.25 &4/5 &0.162 &0.252 & & &$\times$ &\hfil$\circ$ &$\times$ &$\times$ & &VI  &VIII &\cite{L3}\\
2.50 &2/3 &0.155 &0.278 & & & &$\bullet \circ$ &$\times$ &$\times$ &$\bullet$ &II  &IV   &\cite{L3}\\
\hline
\end{tabular}
\label{tab:emmodels}
\end{table}

\begin{table}[htb]
\centering
\caption{Same as Table \ref{tab:emmodels}, but for unequal-mass binaries.}
\begin{tabular}{@{}|cc|cccccc|cc|c|@{}}
\hline
\multicolumn{2}{|c}{Parameters} &\multicolumn{6}{|c|}{$(M/R)_1+(M/R)_2$} &\multicolumn{2}{|c|}{Table} &Ref. \\
\hline
$\Gamma$ &$n$ &$(0.08+0.10)$ &$(0.10+0.12)$ &$(0.12+0.14)$ &$(0.14+0.16)$ &$(0.16+0.18)$ &$(0.18+0.20)$ &Cor &NoS & \\
\hline
1.80 &5/4 &$\times$ &$\times$ &$\times$ & & &  &XI  &XIII &\cite{L3}\\
2.00 &1   & & &$\bullet$ &$\times$ &$\bullet$ &  &II  &IV   &\cite{L2}\\
2.25 &4/5 & & &$\times$ &$\times$ &$\times$ &  &VII &IX   &\cite{L3}\\
2.50 &2/3 & & & &$\times$ &$\times$ &$\times$  &III &V    &\cite{L3}\\
\hline
\end{tabular}
\label{tab:ummodels}
\end{table}

\subsection{PN diagnosis of Meudon data}

The tables of Refs. \cite{L1,L2,L3} display 
the total ADM mass $\bar M$, angular momentum $\bar J$
and angular velocity $\bar \Omega$
of the system (among other variables) for the various models.
Overbars denote that all variables are in polytropic units.
One piece of information missing from those tables 
is the ADM mass of each star in isolation; these
values were kindly provided to us by Keisuke Taniguchi, along with
tables of all data that went into those references, but with full
untruncated precision.  

It is then straightforward to compute $E/m$ and $J/m^2$ using Eqs. 
(\ref{EJtot}) -- (\ref{BCcoeffs})
as functions of $m\Omega$, for each value of $\Gamma$ (which
determines the apsidal constants) and the compactness parameter, together
with the provisional assumption that $e=0$.  From the numerical data
the corresponding binding energy and angular momentum are given by
\begin{eqnarray}
\left (\frac{E}{m} \right )_{\rm Num}  &=& \frac{\bar M}{{\bar M}_{\rm
ADM,0} } -1 \,,
\nonumber \\
\left (\frac{J}{m^2}\right )_{\rm Num}  &=& 
\frac{\bar J}{{\bar M}_{\rm ADM,0}^2} \,,
\nonumber \\
(m\Omega )_{\rm Num} &=& {\bar M}_{\rm ADM,0} {\bar \Omega} \,,
\label{EJnumdefs}
\end{eqnarray}
where
${\bar M}_{\rm ADM,0}$ is the sum of the ADM masses of the two stars
in isolation.  We then plot $(E/m)_{\rm Num}$ and $(J/m^2)_{\rm Num}$ 
against $(m\Omega)_{\rm Num}$.   
Since
the final quantities $E/m$, $J/m^2$  and $m\Omega$ are dimensionless, 
the polytropic units scale out. 
The compactness parameter is quoted as $({M}/{R})$, where here ${M}$ 
is the ADM mass of the isolated star, and $R$ is the
areal, or Schwarzschild  radius (the value is independent of whether
polytropic or normal units are used); as 
a result, using Eq. (\ref{radialcoord}),
we can read off the
value
\begin{equation}
(q)_{\rm Num} = \left (\frac{R}{M}\right ) -1 \,.
\end{equation}

\begin{figure*}[htb]
\begin{center}
\begin{tabular}{cc}
\epsfig{file=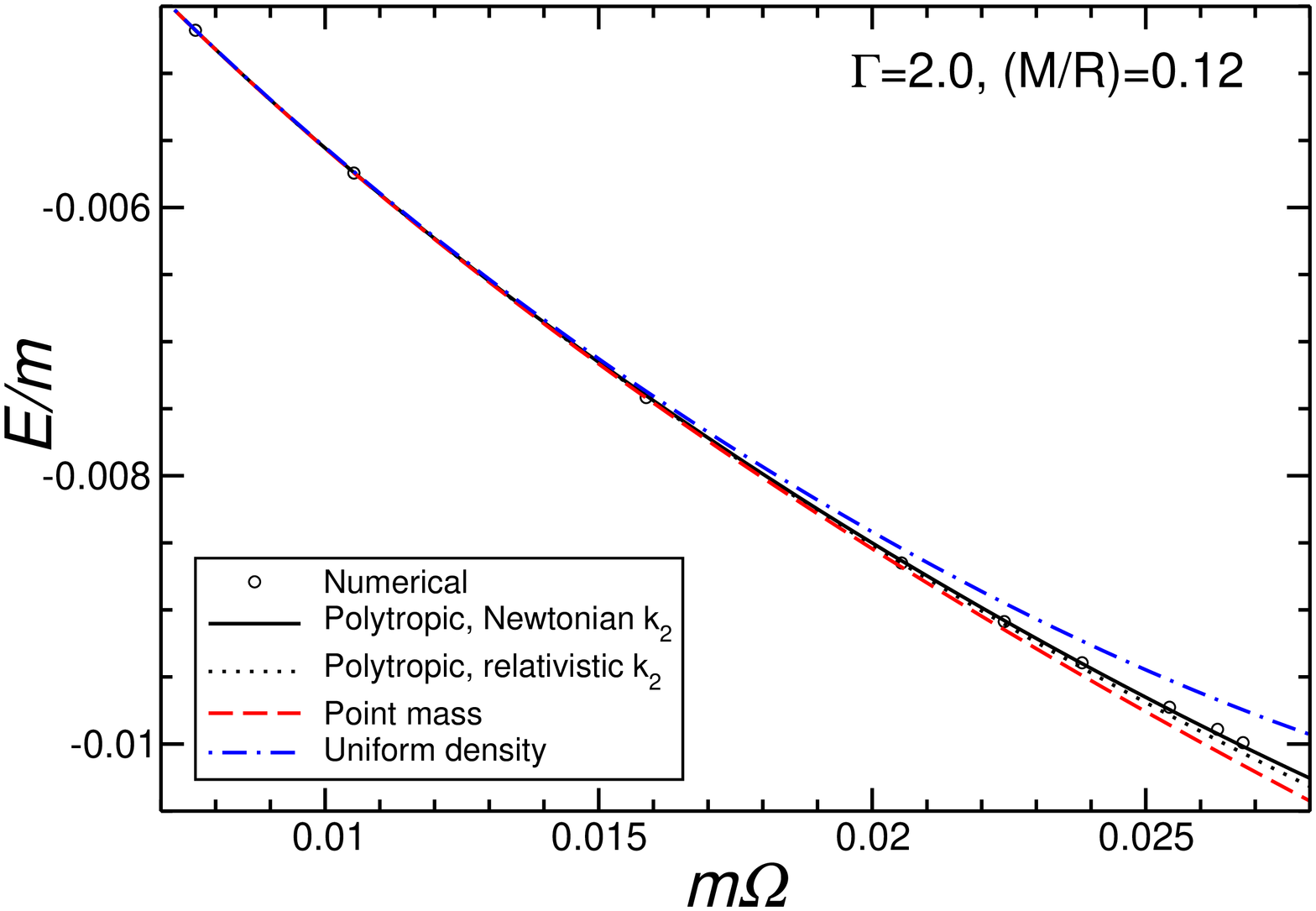,width=6cm,angle=-90} &
\epsfig{file=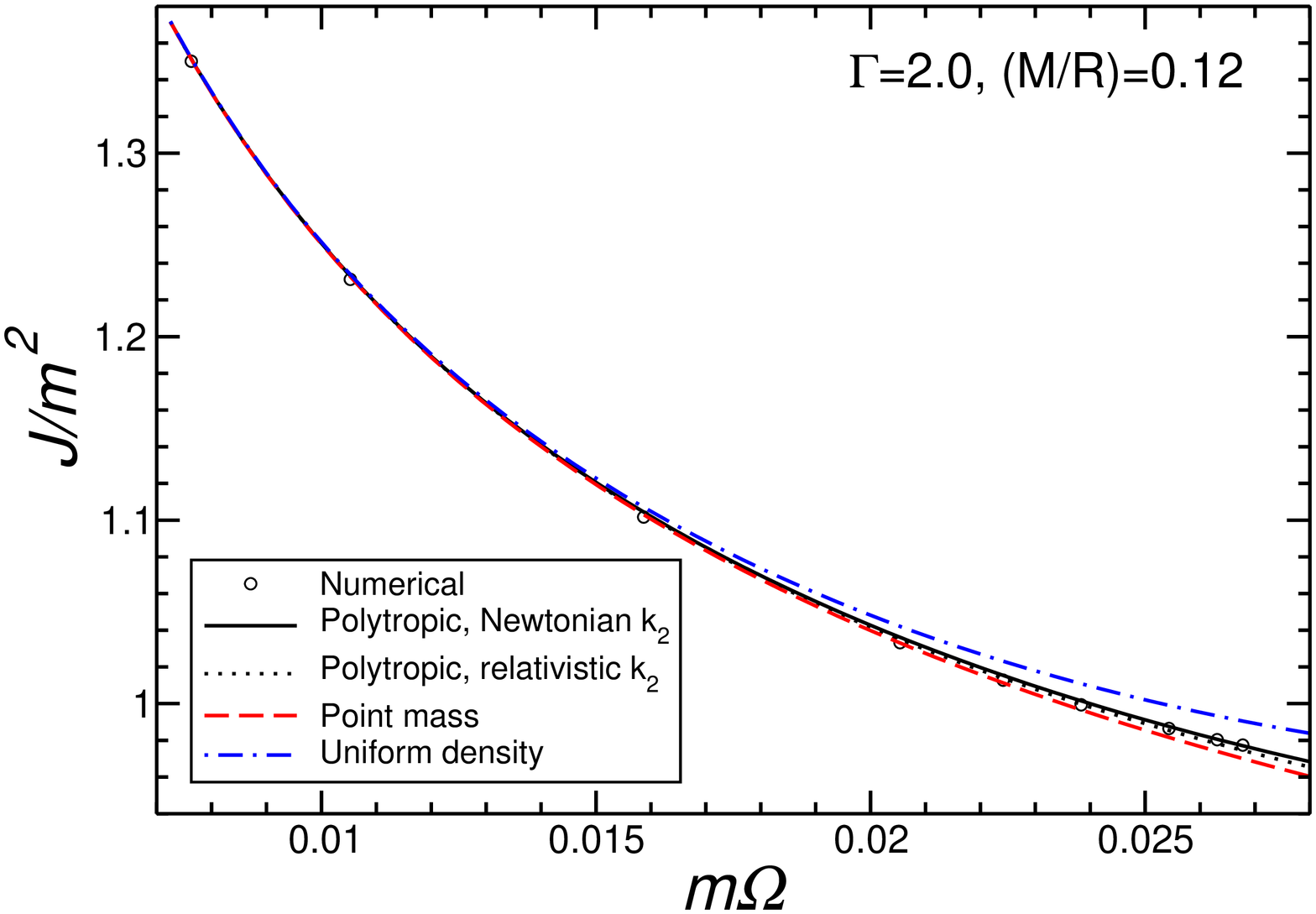,width=6cm,angle=-90} \\
\epsfig{file=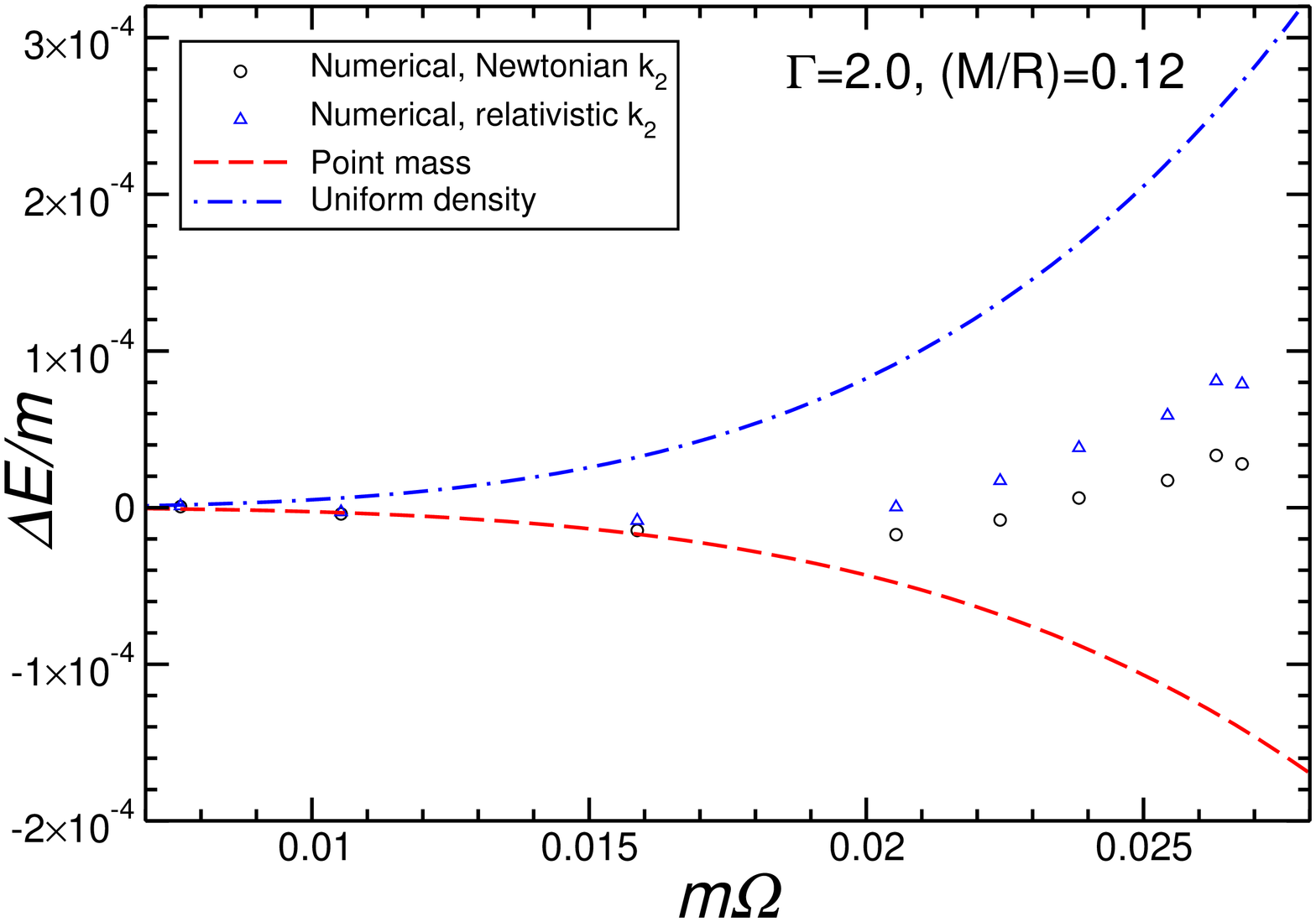,width=6cm,angle=-90} &
\epsfig{file=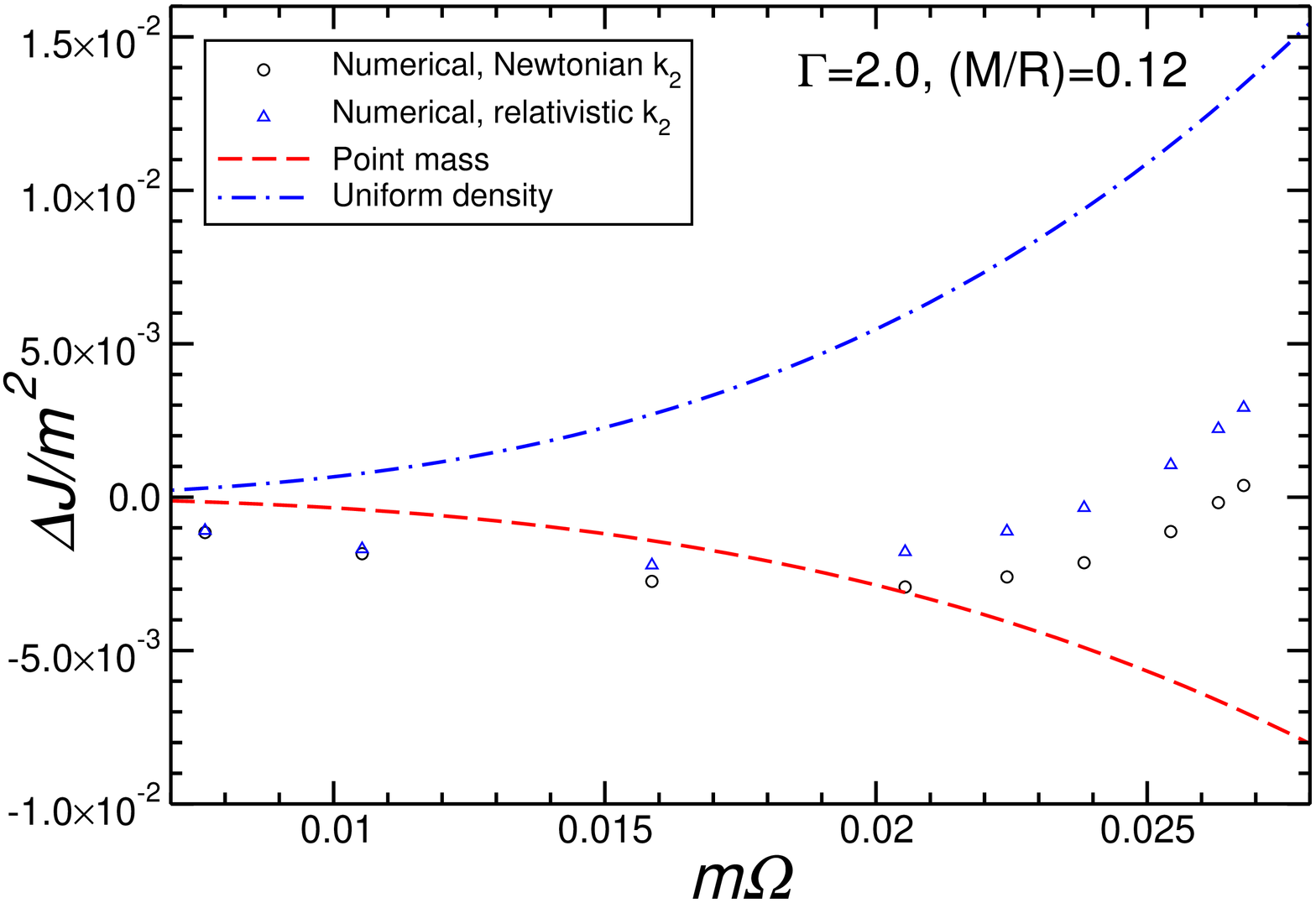,width=6cm,angle=-90} \\
\end{tabular}
\caption{Binding energy and angular momentum vs. $m\Omega$ for
equal masses, $\Gamma=2.0$, $(M/R)_1=(M/R)_2=0.12$. 
Top: 
circles are the
numerical data points, solid (black) line is the PN diagnostic using the
Newtonian apsidal constant values (Table \ref{tab:apsidaln}) for
$\Gamma=2$, dashed (red) line has no tidal contributions; dot-dashed (blue)
line
uses the uniform density values for the apsidal constants; dotted
(black) line uses the  ``relativistically
corrected'' apsidal constants of Eq.~(\ref{k2rel}). 
Bottom: Numerical data (circles), point-mass (dashed-red) and
uniform-density (dot-dashed-blue) plotted relative to $\Gamma=2$ 
PN diagnostic as 
the baseline. Triangles denote numerical data plotted
relative to  $\Gamma=2$
PN diagnostic using the relativistically corrected apsidal constants.
\label{201212-D201212}}
\end{center}
\end{figure*}

\begin{figure*}[htb]
\begin{center}
\begin{tabular}{cc}
\epsfig{file=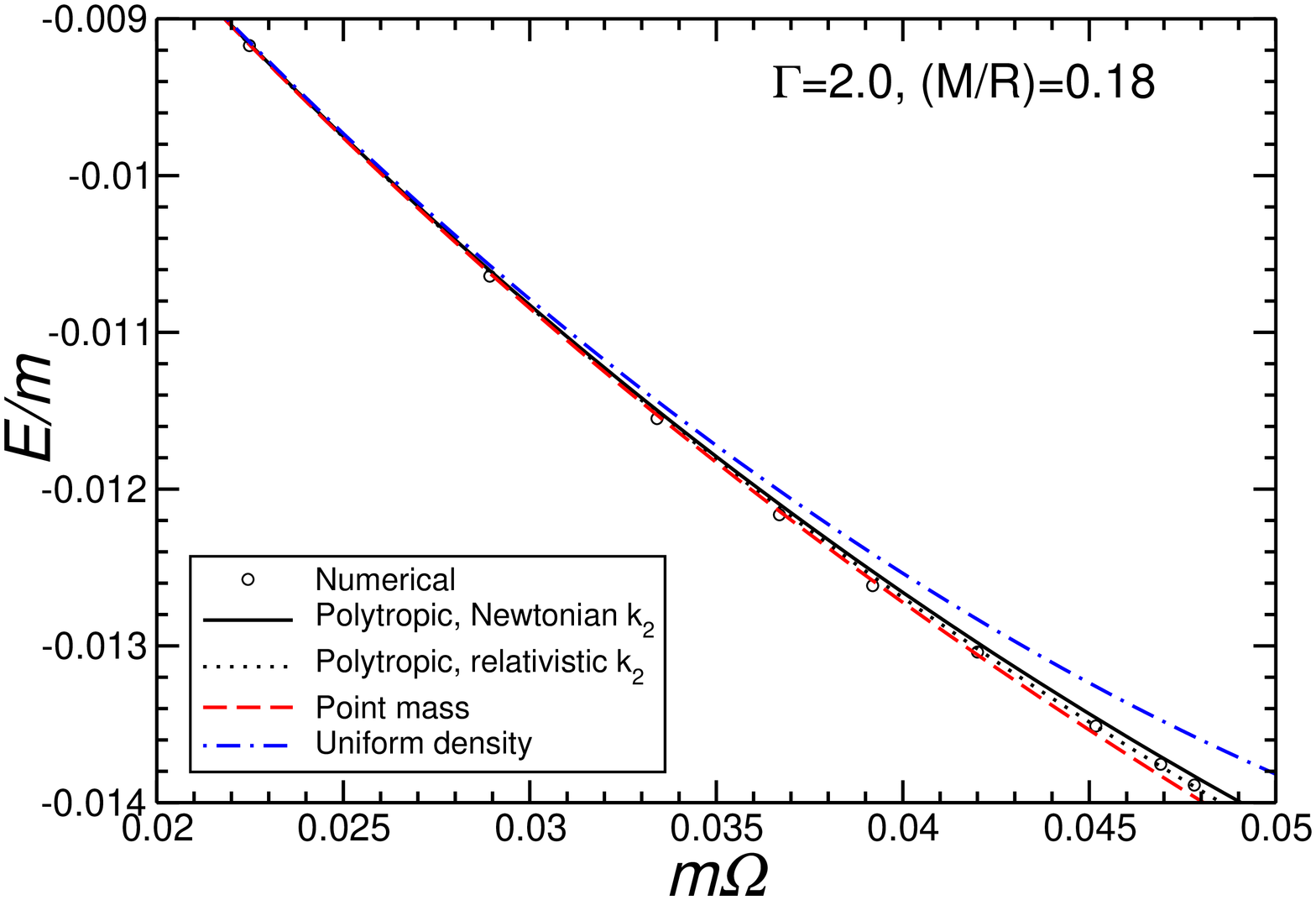,width=6cm,angle=-90} &
\epsfig{file=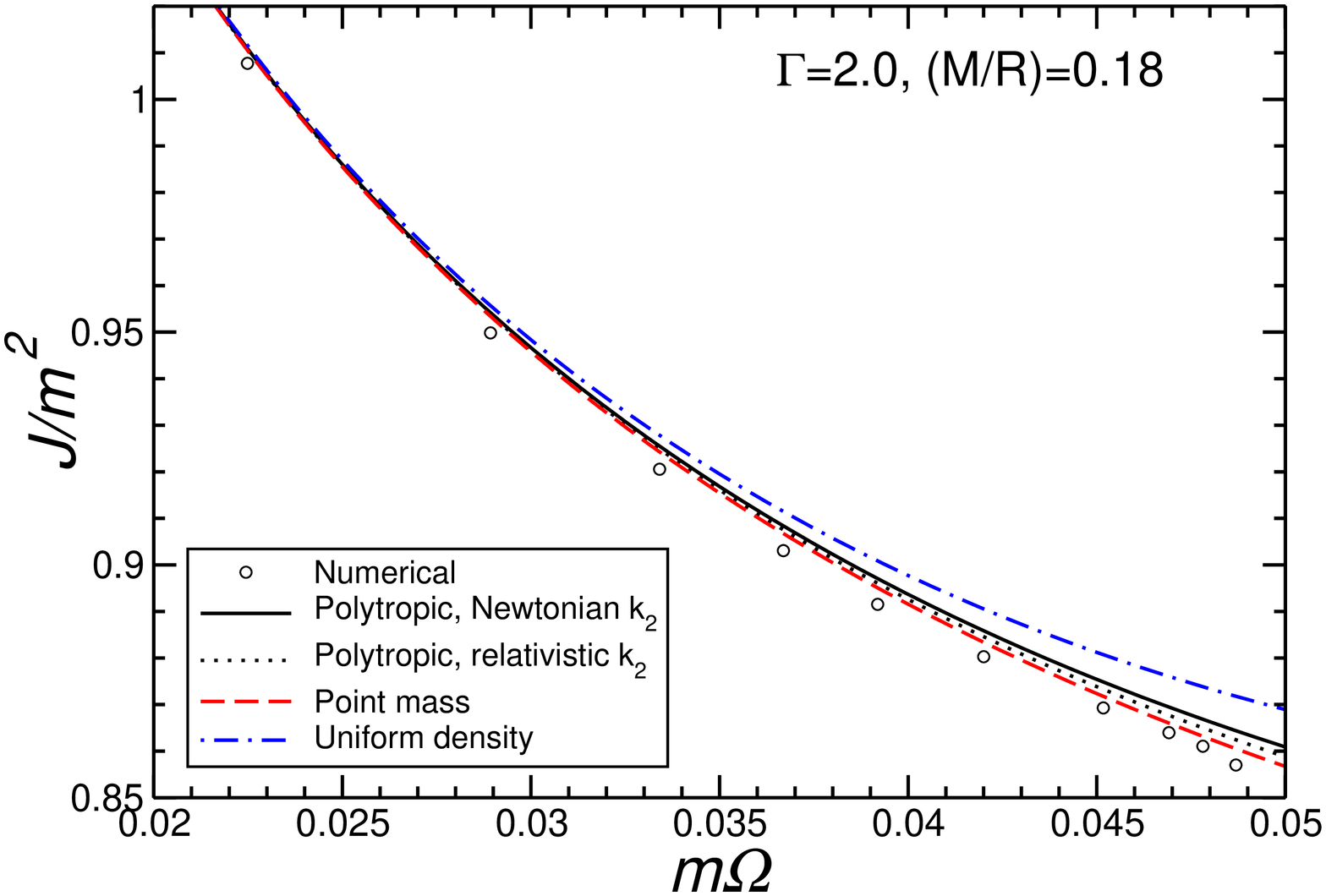,width=6cm,angle=-90} \\
\epsfig{file=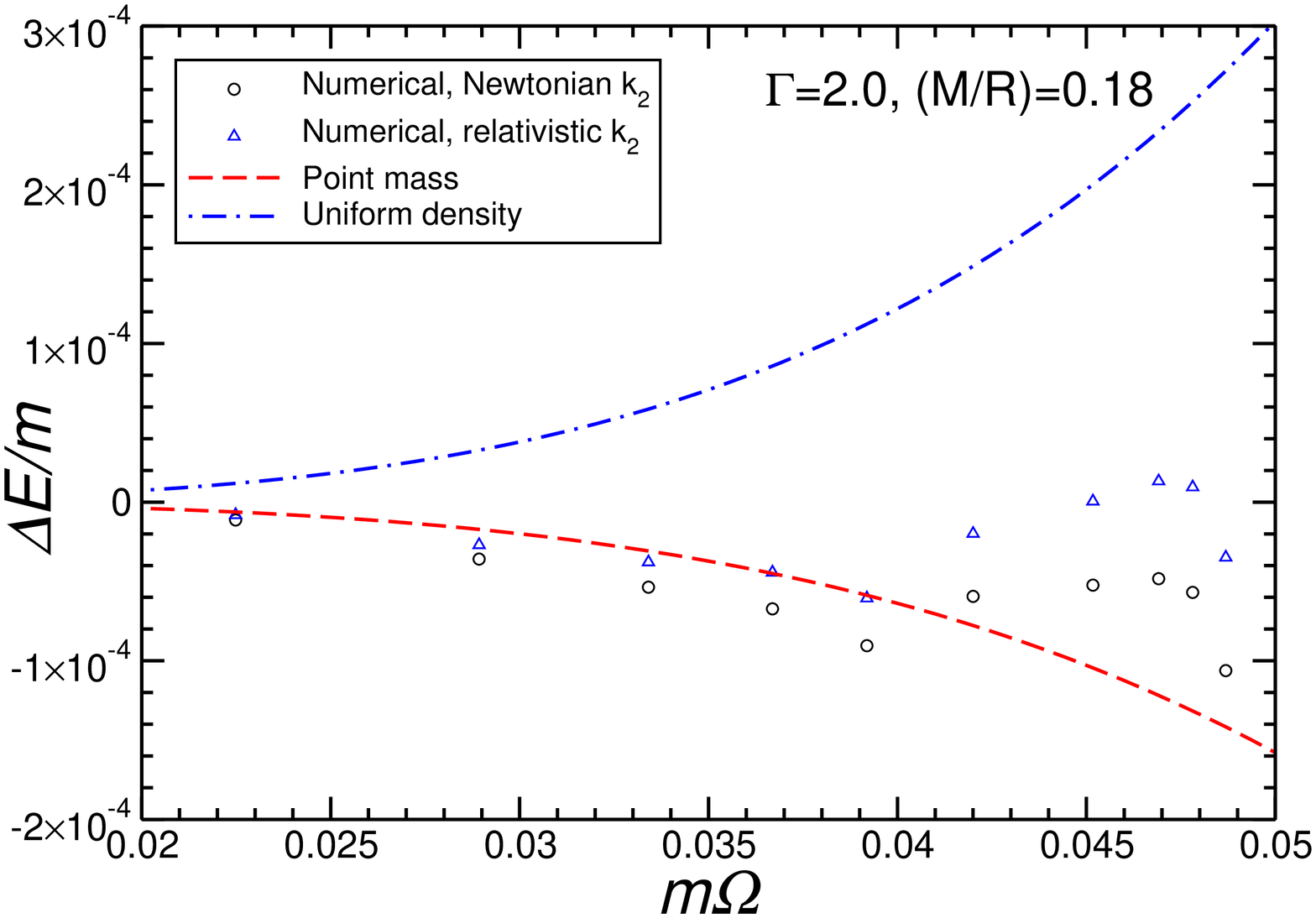,width=6cm,angle=-90} &
\epsfig{file=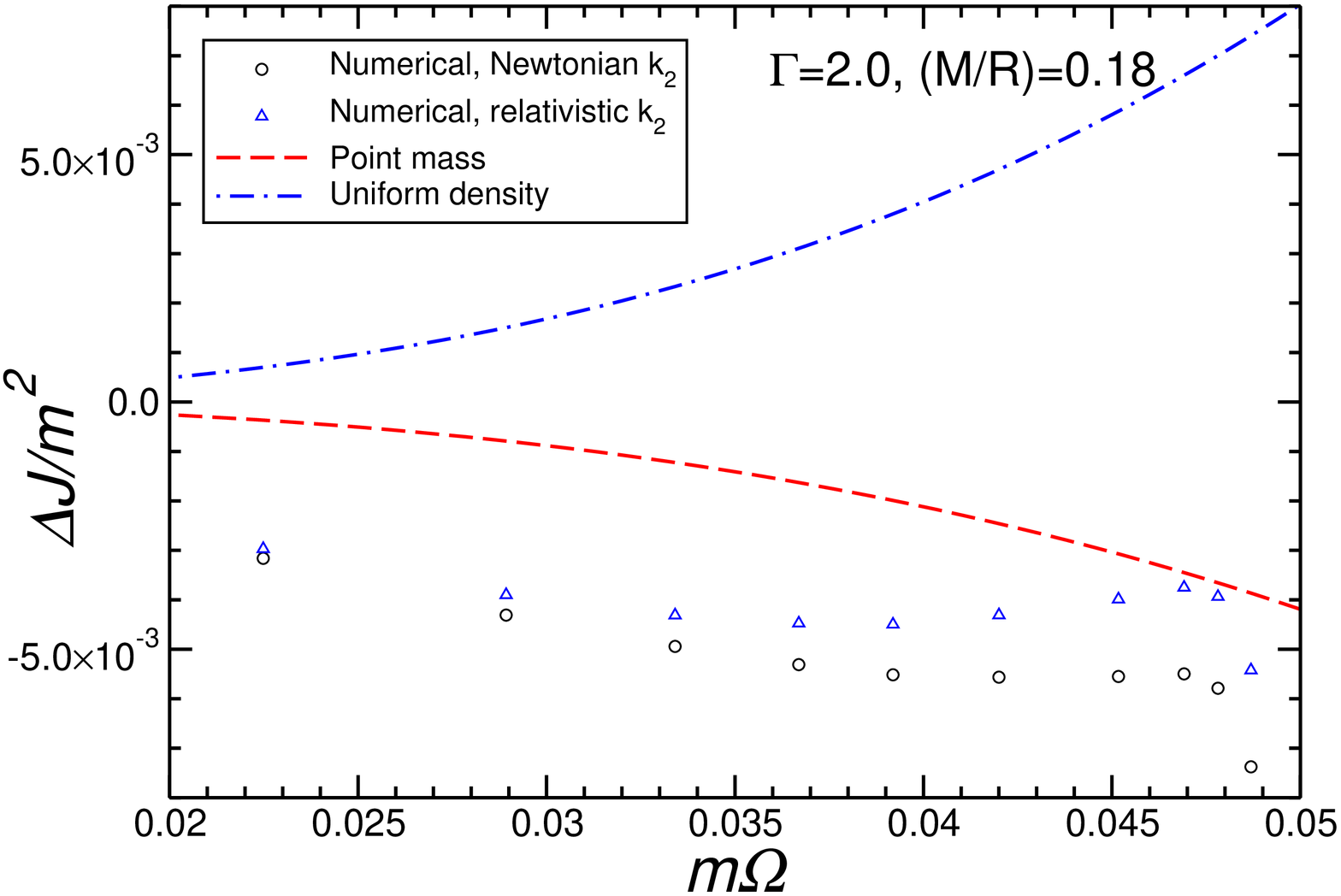,width=6cm,angle=-90} \\
\end{tabular}
\caption{Same as Fig. \ref{201212-D201212}, for 
$\Gamma=2.0$, $(M/R)_1=(M/R)_2=0.18$. \label{201818-D201818}}
\end{center}
\end{figure*}

Figures \ref{201212-D201212} and \ref{201818-D201818} show the
results for equal masses, $\Gamma=2$ and compactness values $0.12$ and $0.18$,
corresponding respectively to neutron star harmonic 
radii $7.3$ and $4.5$ in units
of the stellar mass.  The top
figures show, on a gross scale, the overall agreement between the
PN and numerical results.  Using differences between data sets, the
bottom figures magnify the scale:  the circles and triangles show the
difference between the numerical data and our diagnostic using the
Newtonian and relativistically corrected apsidal constants
respectively.  The dashed and dot-dashed
curves illustrate the importance of taking tidal effects
into account; respectively, they show the difference between the PN values for
energy and angular momentum assuming the Newtonian apsidal constants
and those assuming point masses (no tidal effects)
or those assuming uniform density bodies (maximum tidal effects).  
It is clear that the numerical data agree much better with PN formulae
that include appropriate tidal effects than they do with formulae
that assume either point masses or homogeneous bodies.
The agreement is at the level of a percent or 
better over
the range of $m\Omega$ shown.  However, as in the binary BH case, there
appears to be a systematic offset between the data and the diagnostic
for $J/m^2$.   We note that the use of a relativistically corrected
apsidal constant (triangles) has a small effect on the agreement.  As
discussed in Appendix \ref{sec:rotcorr}, relativistic effects tend to
decrease the apsidal constant $k_2$; this has the effect of raising
the data points on the difference plots compared to the Newtonian
diagnostic baseline.

\begin{figure*}[htb]
\begin{center}
\begin{tabular}{cc}
\epsfig{file=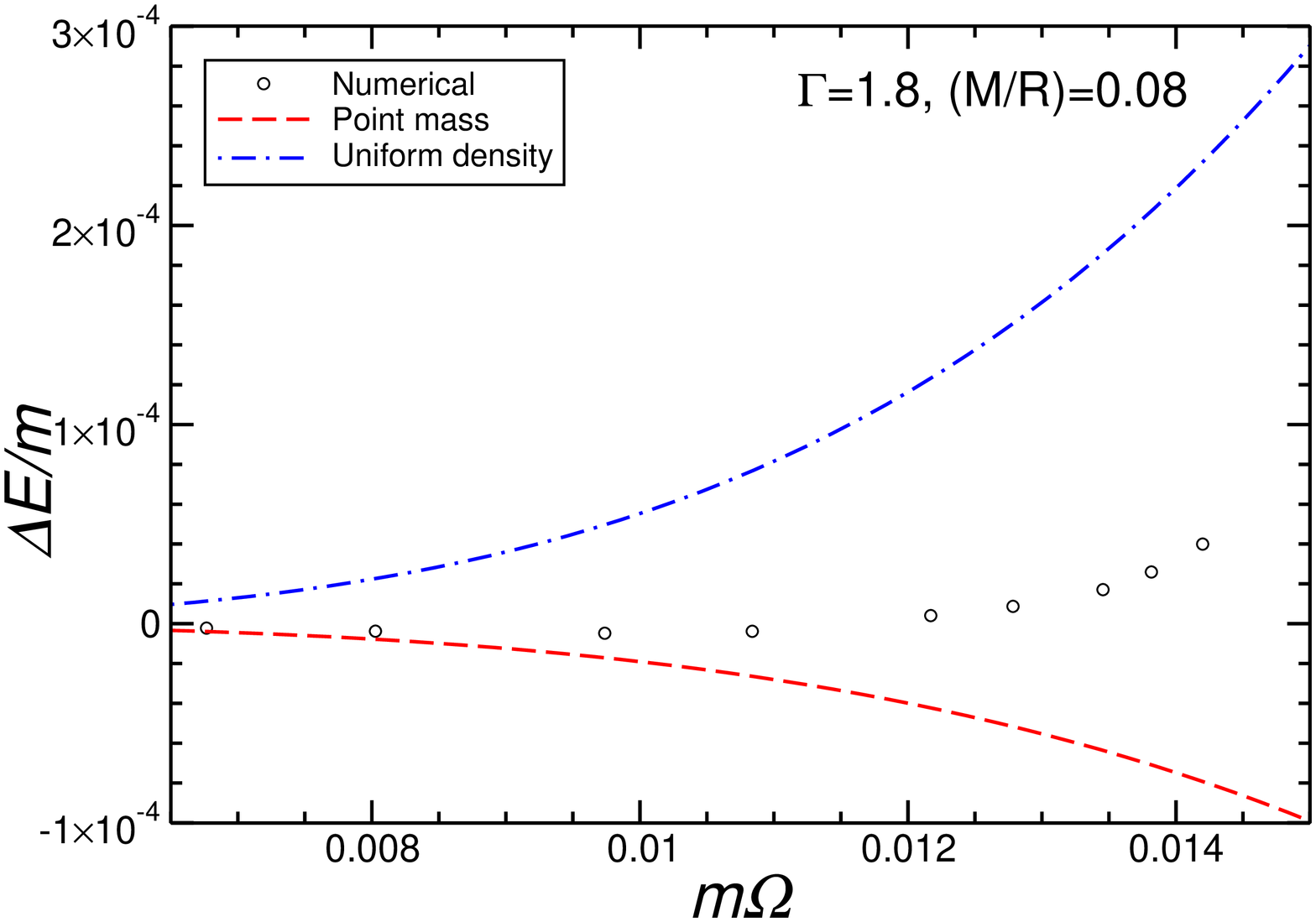,width=6cm,angle=-90} &
\epsfig{file=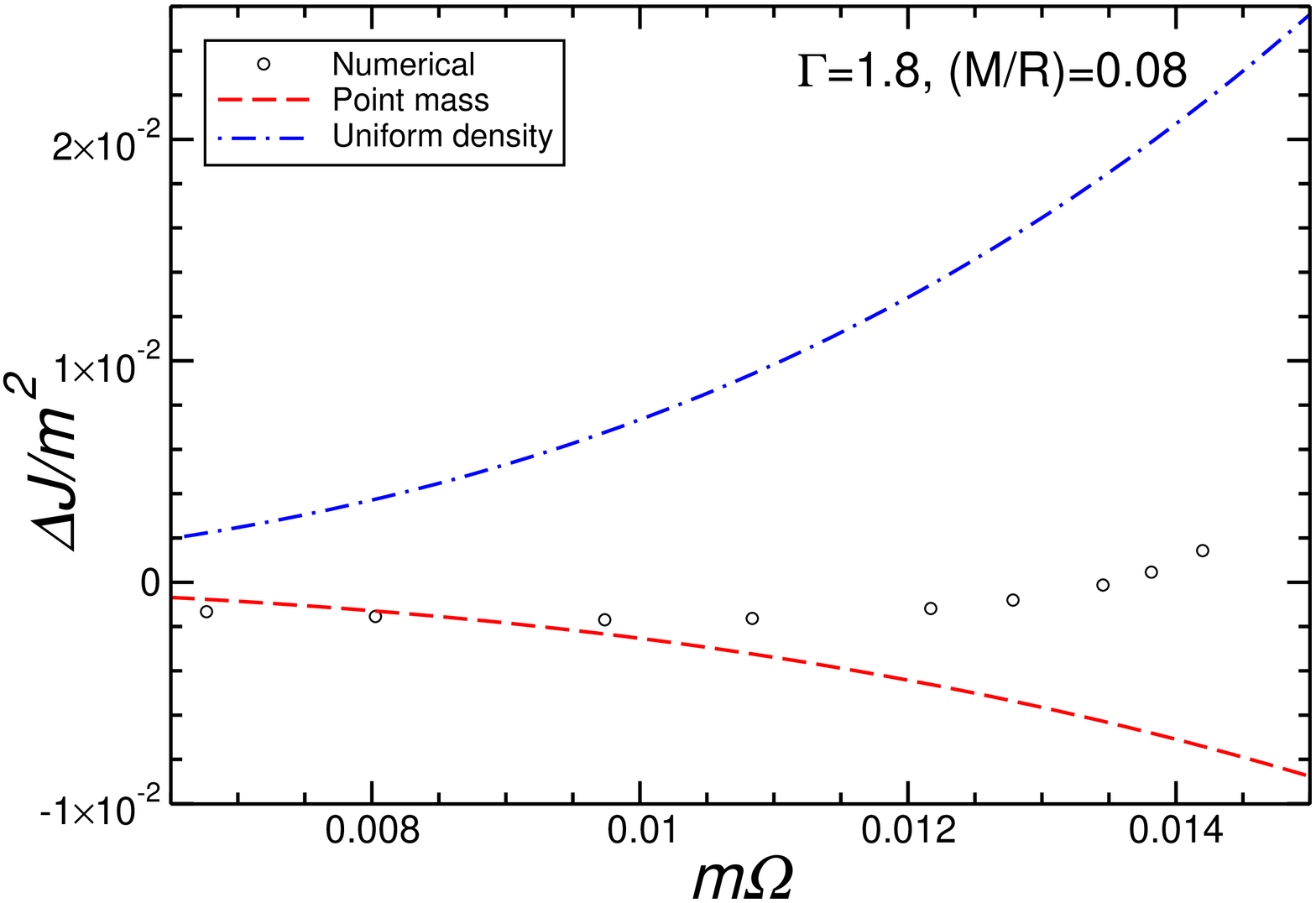,width=6cm,angle=-90} \\
\epsfig{file=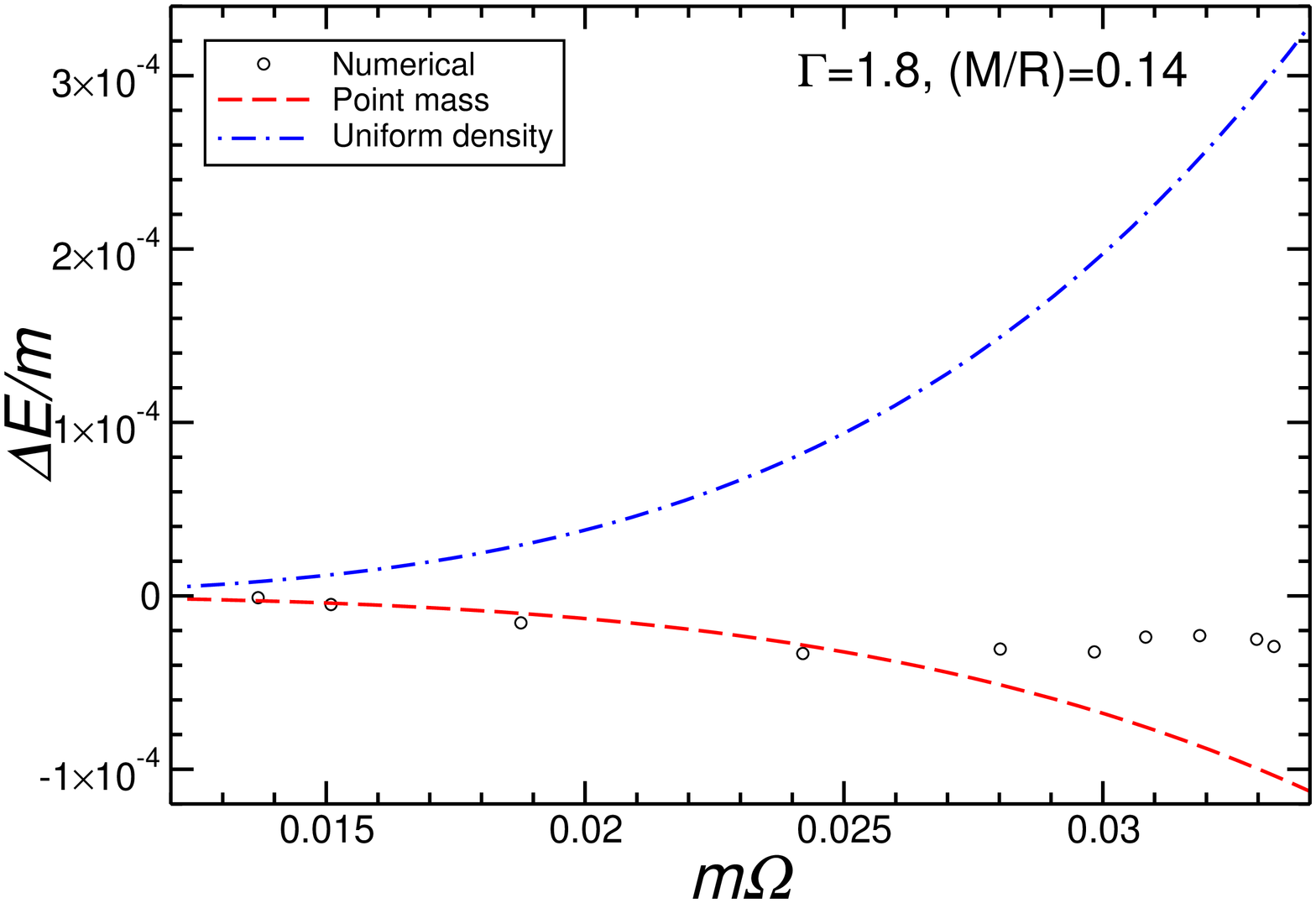,width=6cm,angle=-90} &
\epsfig{file=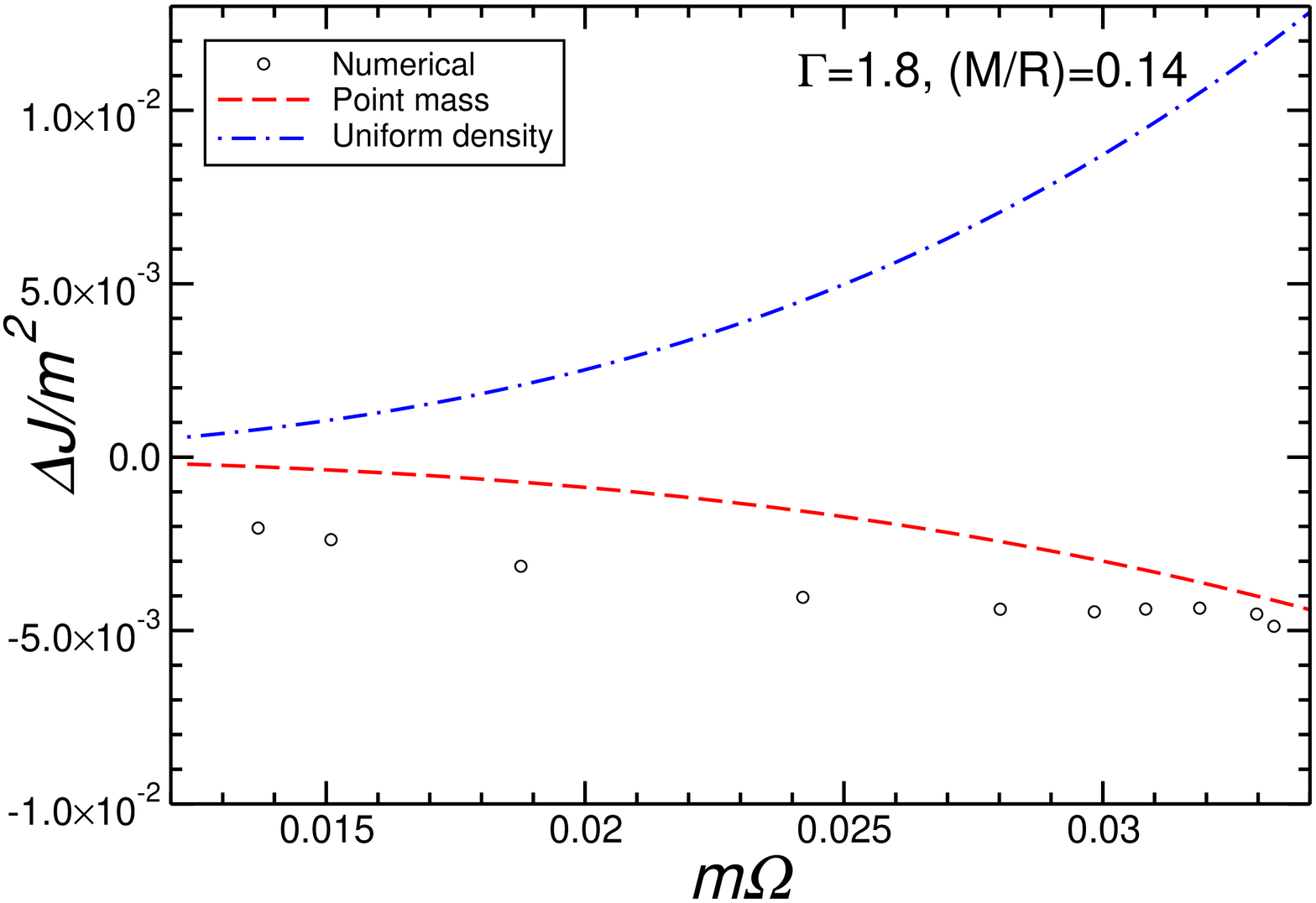,width=6cm,angle=-90} \\
\epsfig{file=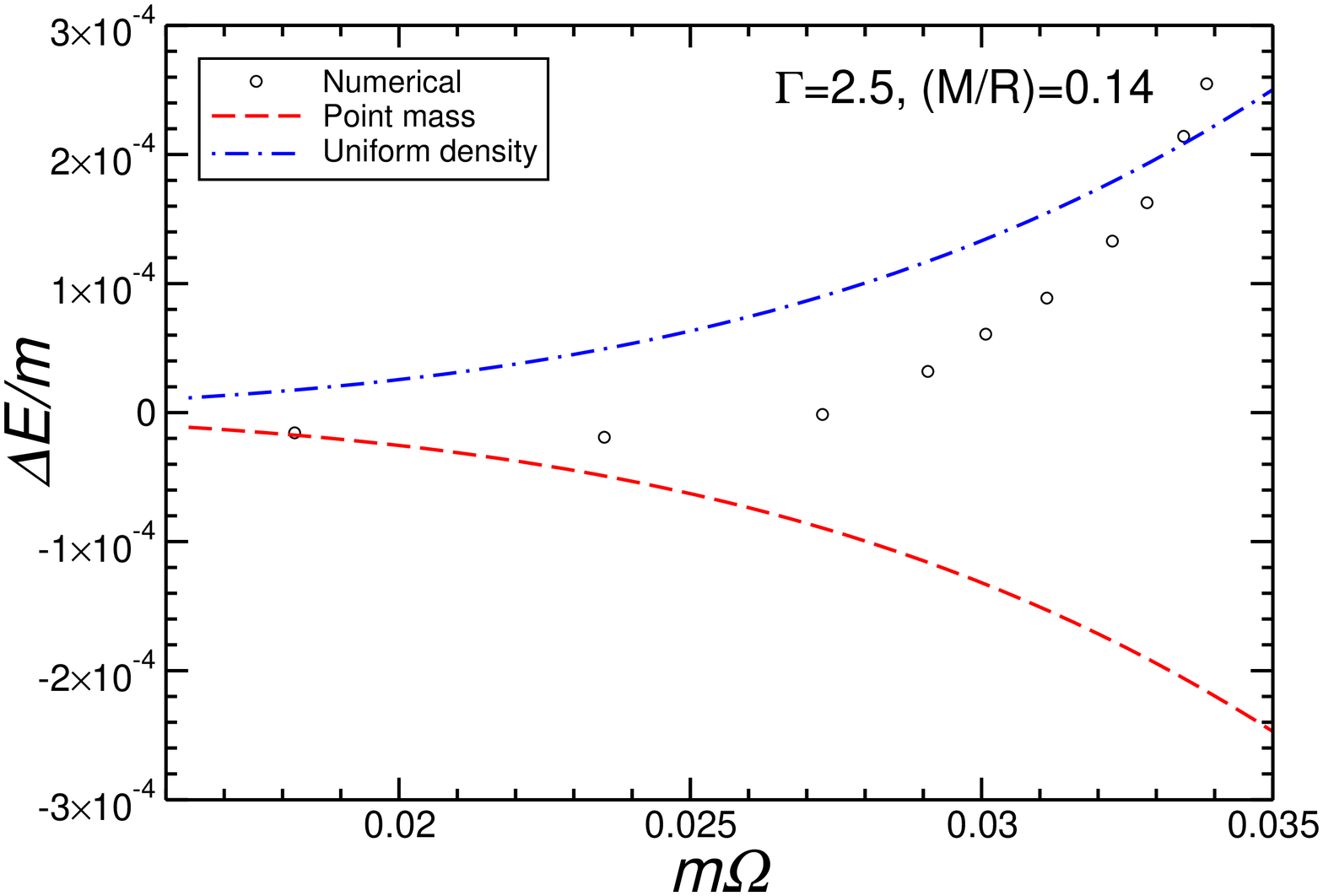,width=6cm,angle=-90} &
\epsfig{file=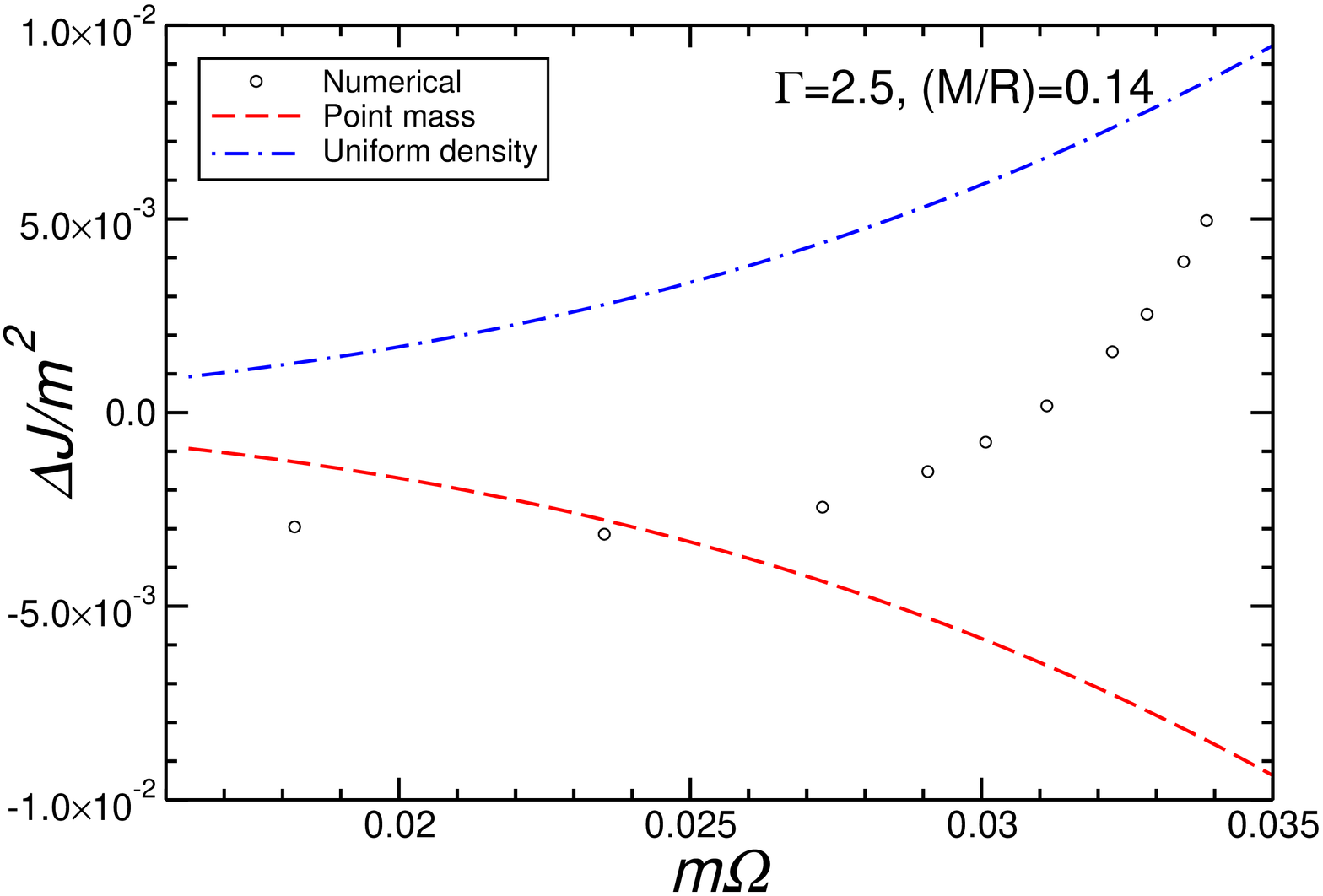,width=6cm,angle=-90} \\
\epsfig{file=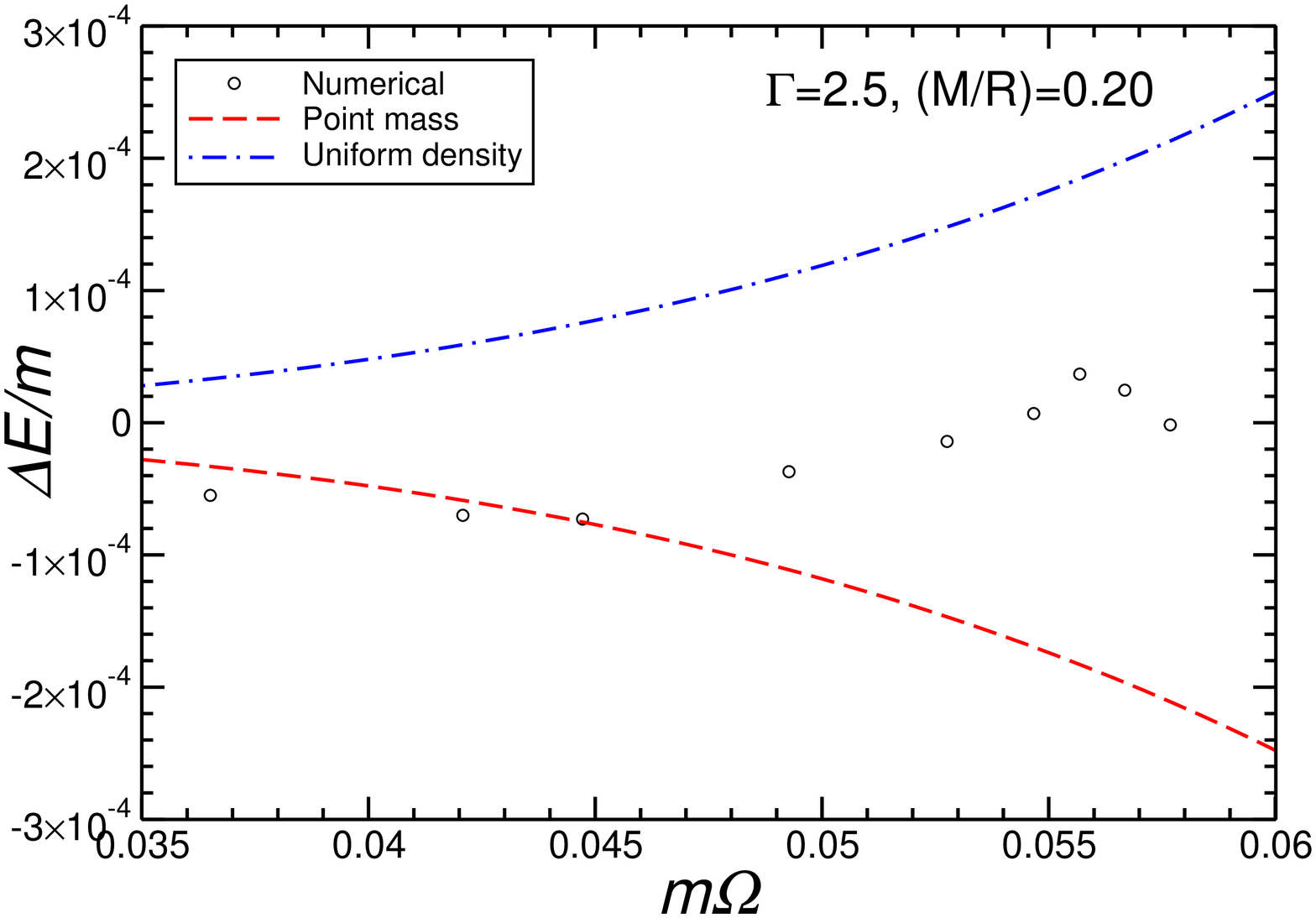,width=6cm,angle=-90} &
\epsfig{file=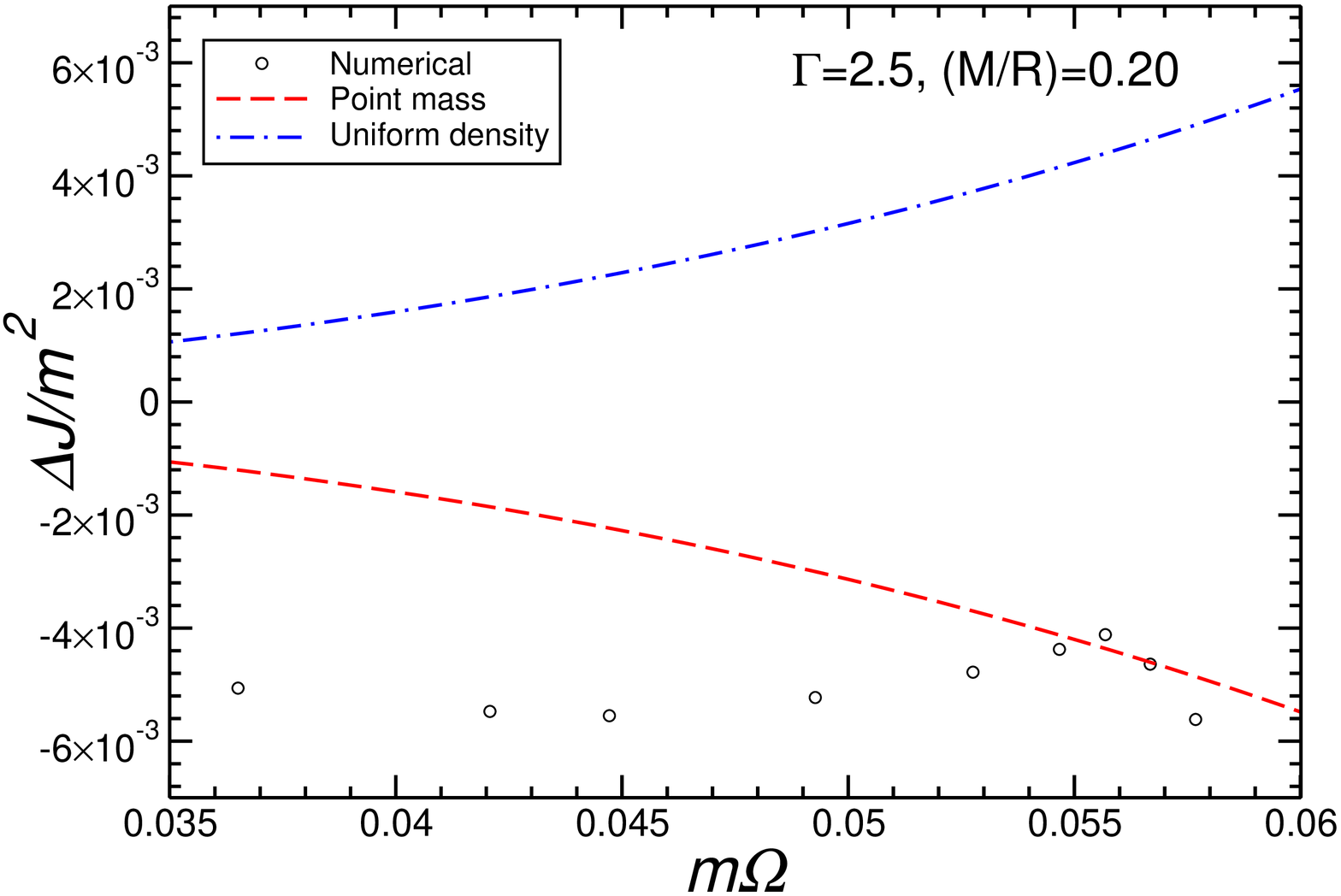,width=6cm,angle=-90} \\
\end{tabular}
\caption{Same as the bottom panels of Fig \ref{201212-D201212}, for various 
values of $\Gamma$ and $M/R$, for equal mass binaries. \label{D180808-D252020}}
\end{center}
\end{figure*}

\begin{figure*}[htb]
\begin{center}
\begin{tabular}{cc}
\epsfig{file=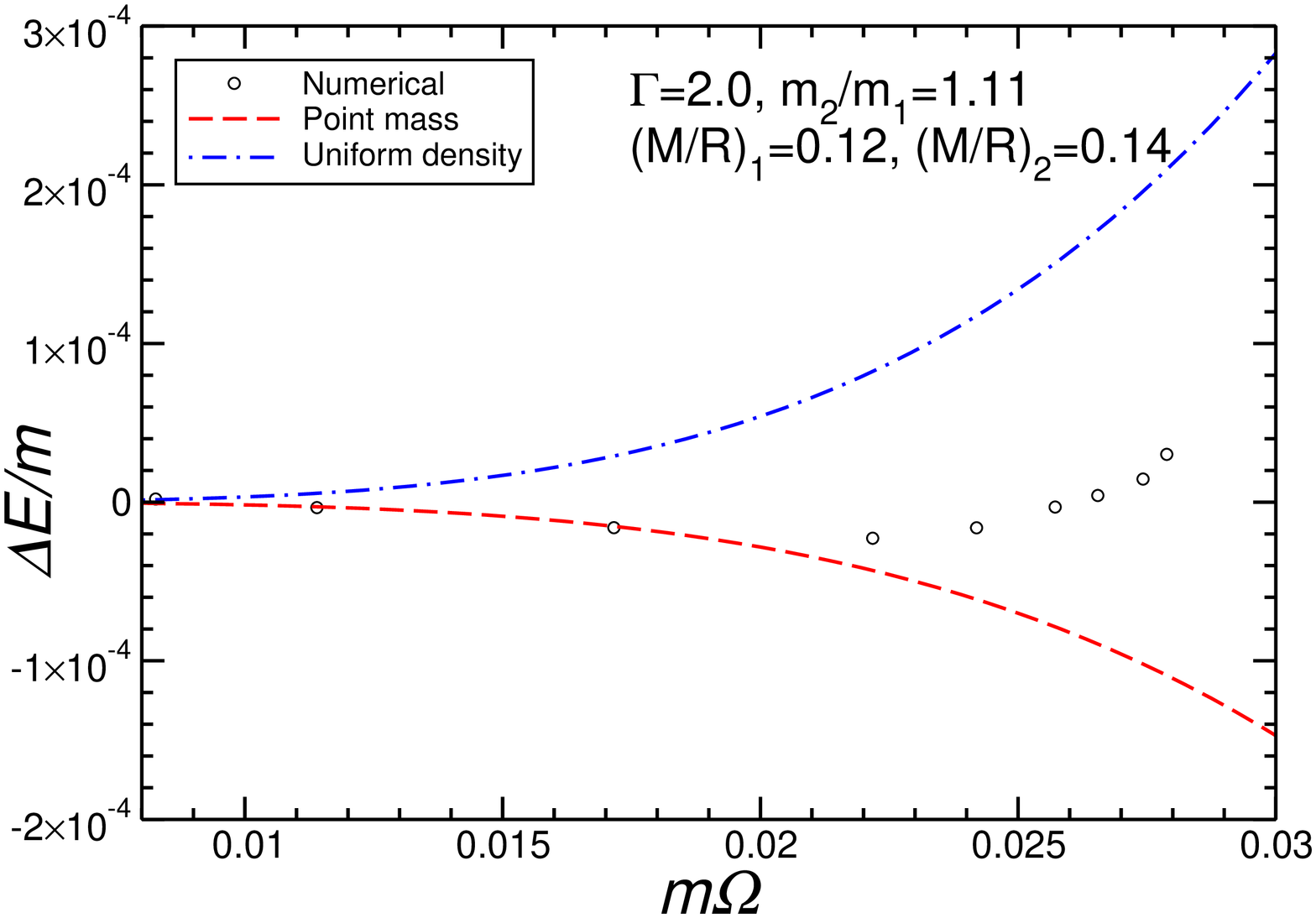,width=6cm,angle=-90} &
\epsfig{file=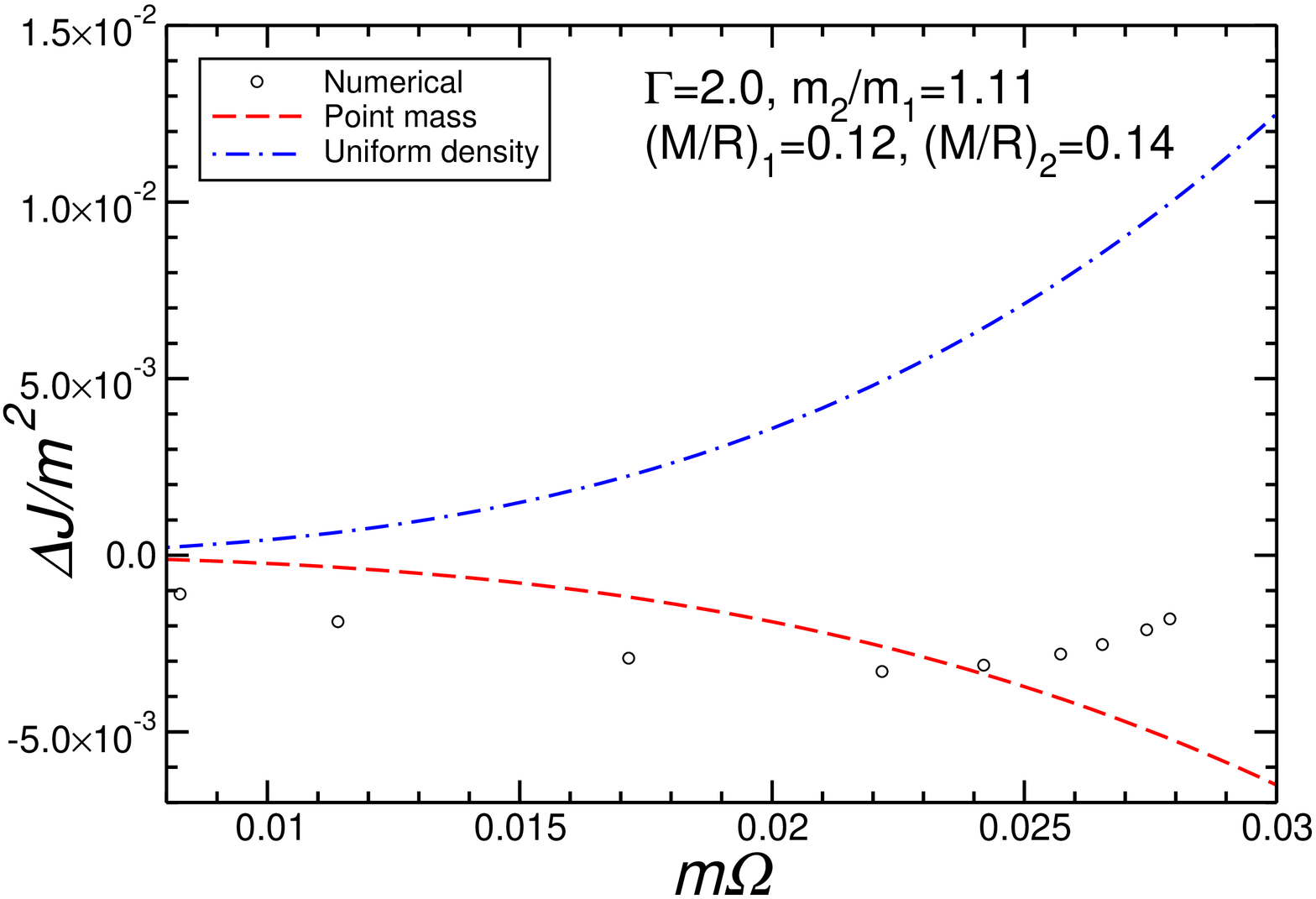,width=6cm,angle=-90} \\
\epsfig{file=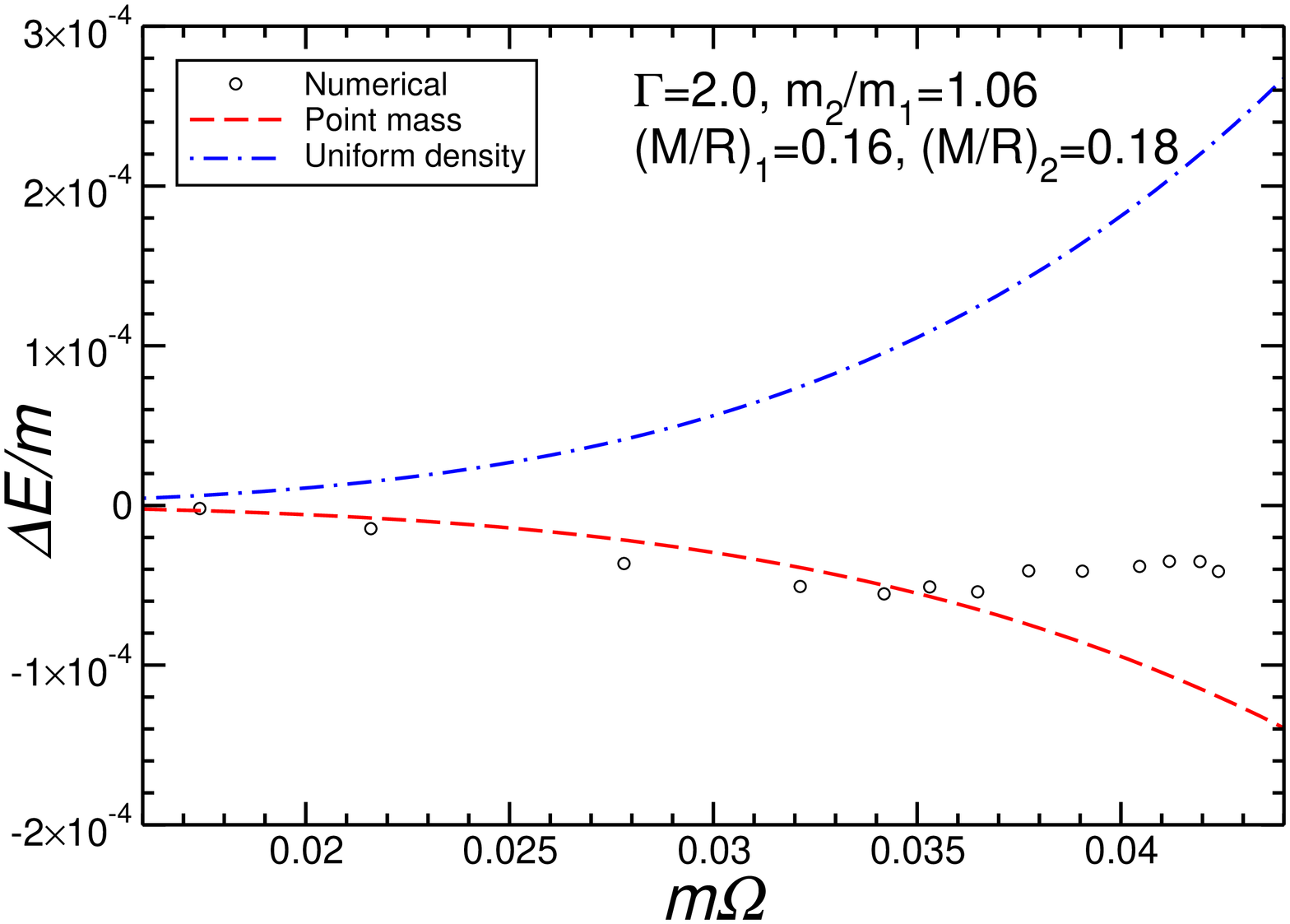,width=6cm,angle=-90} &
\epsfig{file=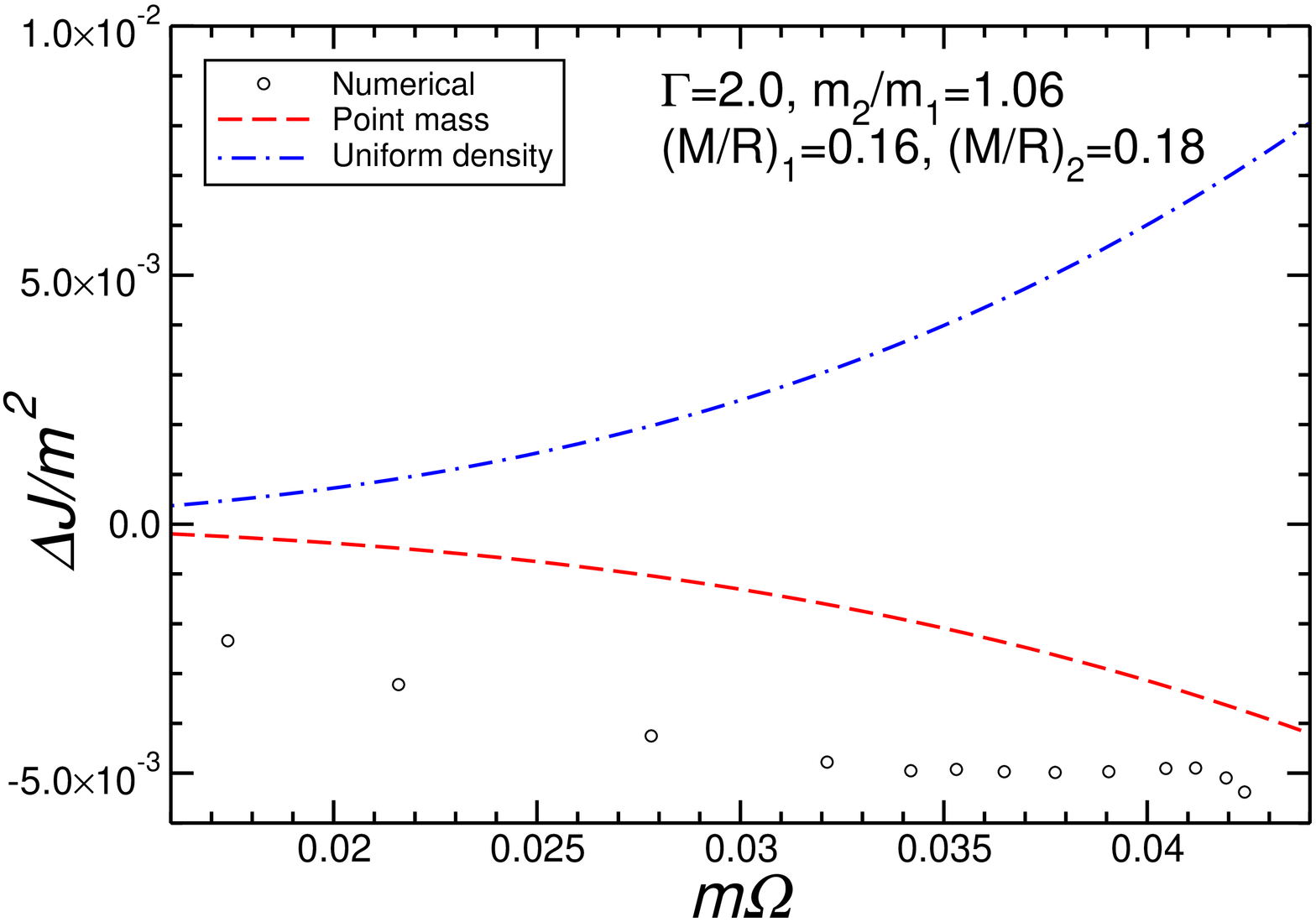,width=6cm,angle=-90} \\
\end{tabular}
\caption{Same as Fig. \ref{D180808-D252020}, 
but for binaries with different masses.
  \label{D201214-D201618}}
\end{center}
\end{figure*}

Figure \ref{D180808-D252020} shows differences relative to the PN
diagnostic baseline for other values of $\Gamma$ and $M/R$ for equal
masses.  The agreement between diagnostic and data seems to worsen as
the EOS progress from soft to hard and the compactness increases; again
there is a systematic offset in $J/m^2$, even in the non-relativistic
limit.  Figure \ref{D201214-D201618} shows differences for selected
configurations of unequal-mass neutron stars.

Instead of comparing the data with a circular-orbit diagnostic, we now
let the eccentricity be a free parameter, and find the value of $e$
that gives the best fit to the data.  Using both the Newtonian and the
relativistic apsidal constants, we show the results for a range of EOS
and compactness parameters in Fig. \ref{fig:eccentricity}.  We note
that the inferred eccentricities are generally smaller than those
inferred from the binary BH initial data, and that there is again a
difference between the values inferred from angular momentum and those
inferred from energy.   The energy-inferred eccentricities are
generally less than 0.005 for all but the high-frequency ends of each
quasiequilibrium sequence.  The end points of these sequences
correspond to binary separations only 25 to 40 percent larger than the
sum of the stellar radii;  at these separations, tidal effects are
sufficiently strong that cusps in the surface shape of the stars are
about to form, signalling the onset of mass shedding.  Not only is our
simple tidal model likely to be inadequate, but also there are larger
errors in the  numerical data near these end points
(see \cite{Taniguchi:2007xm} for discussion).  
 
\begin{figure*}[htb]
\begin{center}
\epsfig{file=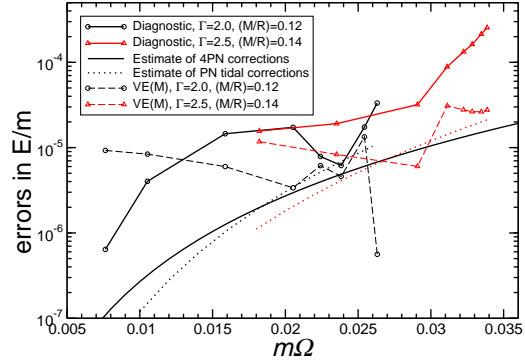,width=6cm,angle=-90}
\caption{Comparison of PN diagnostic eccentricities or errors with
truncation errors in the PN approximation and with ``virial'' errors
in the numerical simulations.
  \label{fig:virial}}
\end{center}
\end{figure*}

The inferred eccentricities are very small, and it is worth asking how
significant they are, given that the PN approximation used in our
diagnostic has truncation errors, and that the numerical simulations
have numerical errors.  Figure \ref{fig:virial} is an attempt to
quantify this.   We plot the energy-inferred eccentricities  for two
representative
models, with $\Gamma=2$, $(M/R)=0.12$, and $\Gamma=2.5$, $(M/R)=0.14$.
The solid curve is ten times what one would crudely estimate for the
4PN contributions to the energy for equal masses, or $10 \times \zeta^5/8 =
1.25 (m\Omega)^{10/3}$, while the dotted curves are $(3/64) q^5 k_2
(m\Omega)^{14/3}$ for the two corresponding values of $q$ and $k_2$,
which represent a post-Newtonian-tidal correction, or $\zeta$ times
the leading tidal term in Eq. (\ref{ETRorbit}).  Also plotted as dashed
curves with data points are the corresponding ``virial'' errors $VE(M)$,
computed from the numerical data, and tabulated in \cite{L2,L3}, where
$VE(M) \equiv |M_{\rm ADM} -M_{\rm Komar}|/M_{\rm ADM}|$, where $M_{\rm
Komar}$ is a certain global integral known as the Komar mass.  This
and similar virial theorems are one indication of the accuracy of the
numerical solution.  We see that PN truncation errors are smaller by
an order of magnitude than the inferred eccentricities or ``errors''
from the diagnostic,
and that the numerical errors are generally smaller, except in the
extreme Newtonian limit.  This gives some confidence that our PN
diagnostic can in many cases ``diagnose'' real physics in the
numerical simulations.

\section{Quasiequilibrium calculations of neutron star-black hole systems}
\label{sec:QENSBH}

The first numerical calculations of fully relativistic neutron
star-black hole binaries were carried out by the Illinois group, for
corotating stars, and in
the limit where the black hole was 10 times more massive than the
neutron star~\cite{baumgarteBHNS}.  However, following a number of
technical improvements, the group recently presented results for
quasiequilibrium sequences of non-spinning NS-BH binaries with a range of   
five mass ratios (including equal masses), and two values of the compactness 
parameter of the neutron star~\cite{Taniguchi:2006yt,Taniguchi:2007xm}.  
They assume a conformally flat spatial metric and solve the
initial value equations for the metric variables using 
multidomain spectral methods.  
Since then, they have also implemented a more
accurate condition for ensuring that the black hole is 
nonspinning, first introduced in
Ref.~\cite{Caudill:2006hw}.  Taniguchi has provided us with
preliminary data for the improved sequences \cite{Taniguchi:PC}.
Using similar methods,
Grandcl\'ement~\cite{Grandclement:2006ht} also obtained
quasiequilibrium NS-BH sequences, for a mass ratio of five and for four
values of the compactness parameter.
To date, all NS-BH calculations have assumed $\Gamma=2$.

As in the NS-NS case, results are given in polytropic units, with
$E/m$, $J/m^2$ and $m\Omega$ already tabulated
as in Eq.~(\ref{EJnumdefs}) 
along with the mass ratio $M_{\rm irr}^{\rm
BH}/M_{\rm ADM,0}^{\rm NS}$ and the ADM mass ${\bar M}_{\rm ADM,0}^{\rm NS}$
and compactness parameter $M_{\rm
ADM,0}^{\rm NS}/R_0$ of the isolated neutron star, 
where $R_0$ is again the areal radius.  
From these data, we compute 
the factor $q$ for the neutron star from 
$q = R_0/M_{\rm ADM,0}^{\rm NS} -1$.  

\begin{figure*}[htb]
\begin{center}
\begin{tabular}{cc}
\epsfig{file=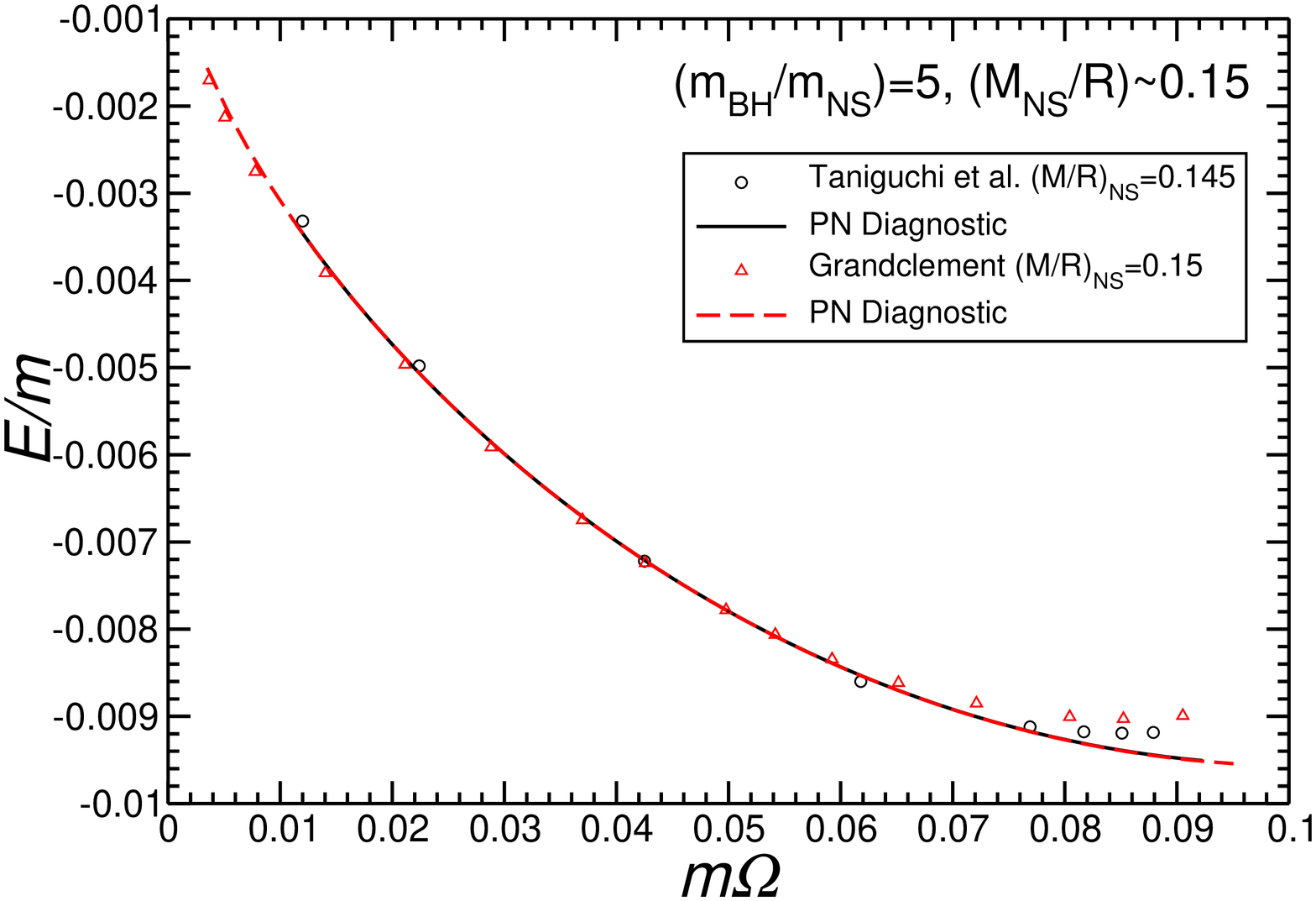,width=6cm,angle=-90} &
\epsfig{file=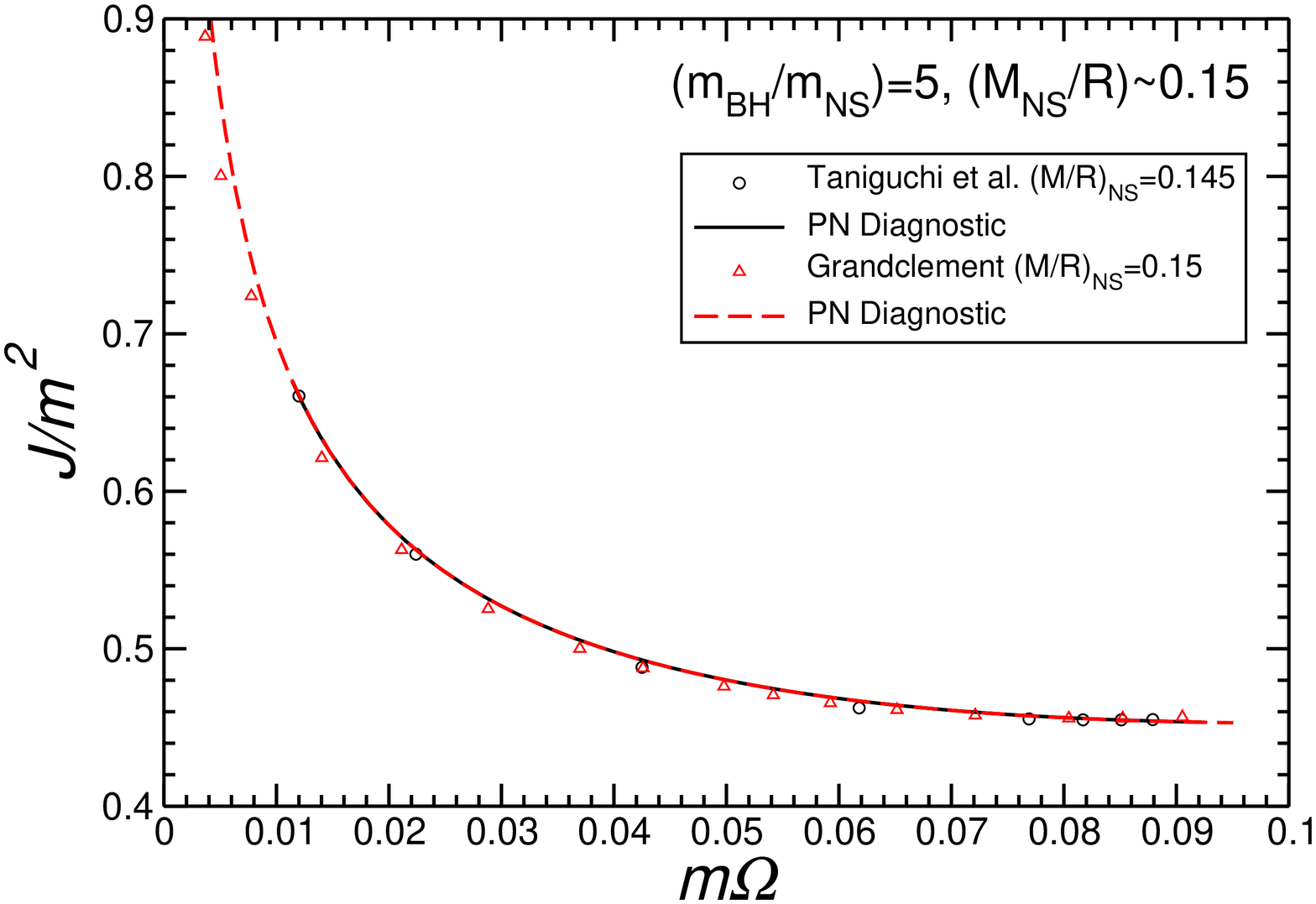,width=6cm,angle=-90} \\
\end{tabular}
\caption{Binding energy and angular momentum 
vs. $m\Omega$
for a black hole-neutron star binary with
mass ratio 5.  Plotted are data from 
  Taniguchi {\em et al.} (circles) and from
Grandcl\'ement (triangles), along with the corresponding PN diagnostic
curve for circular orbits. }
\label{fig:EJ-nsbh}
\end{center}
\end{figure*}

\begin{figure*}[htb]
\begin{center}
\begin{tabular}{cc}
\epsfig{file=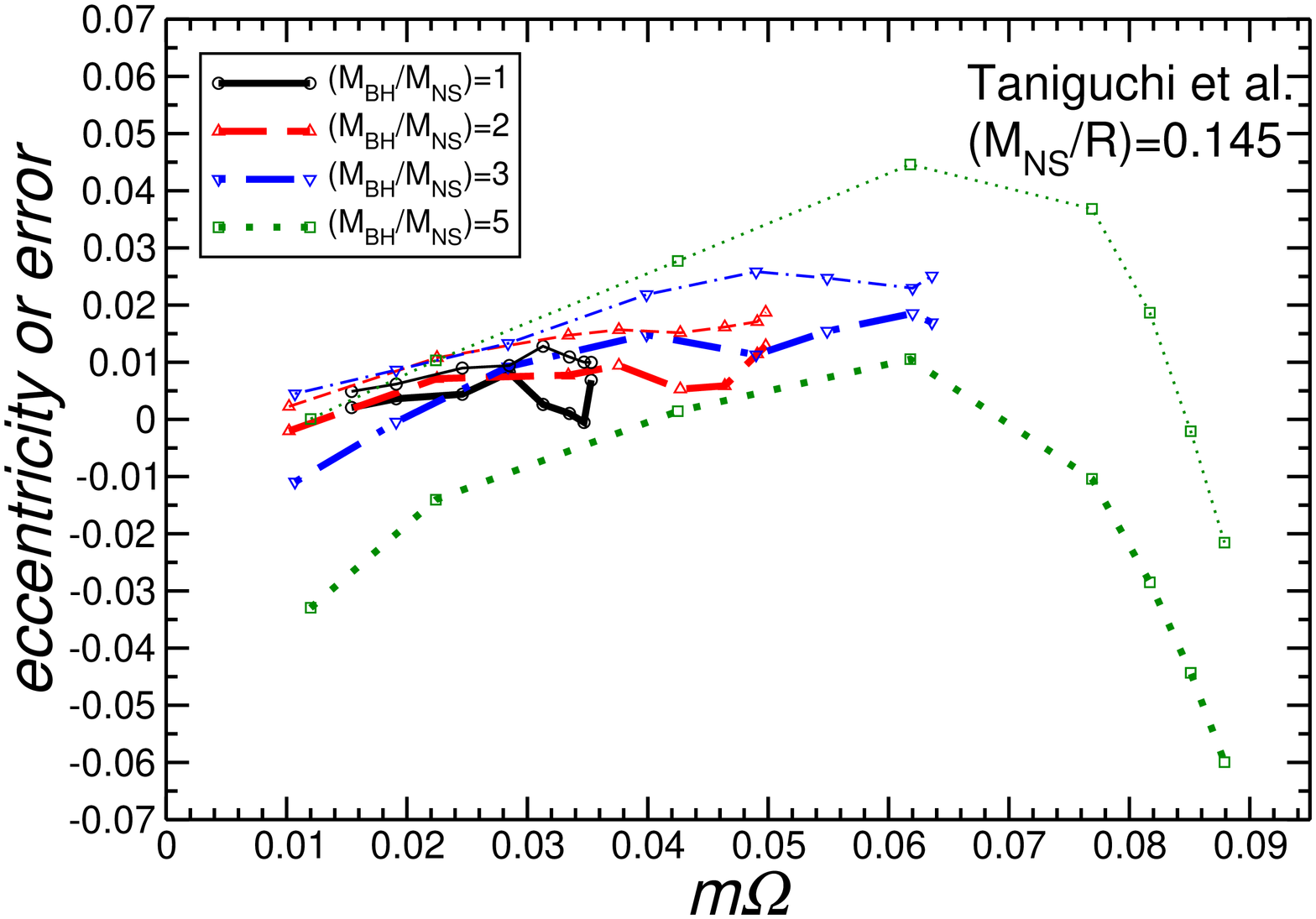,width=6cm,angle=-90} &
\epsfig{file=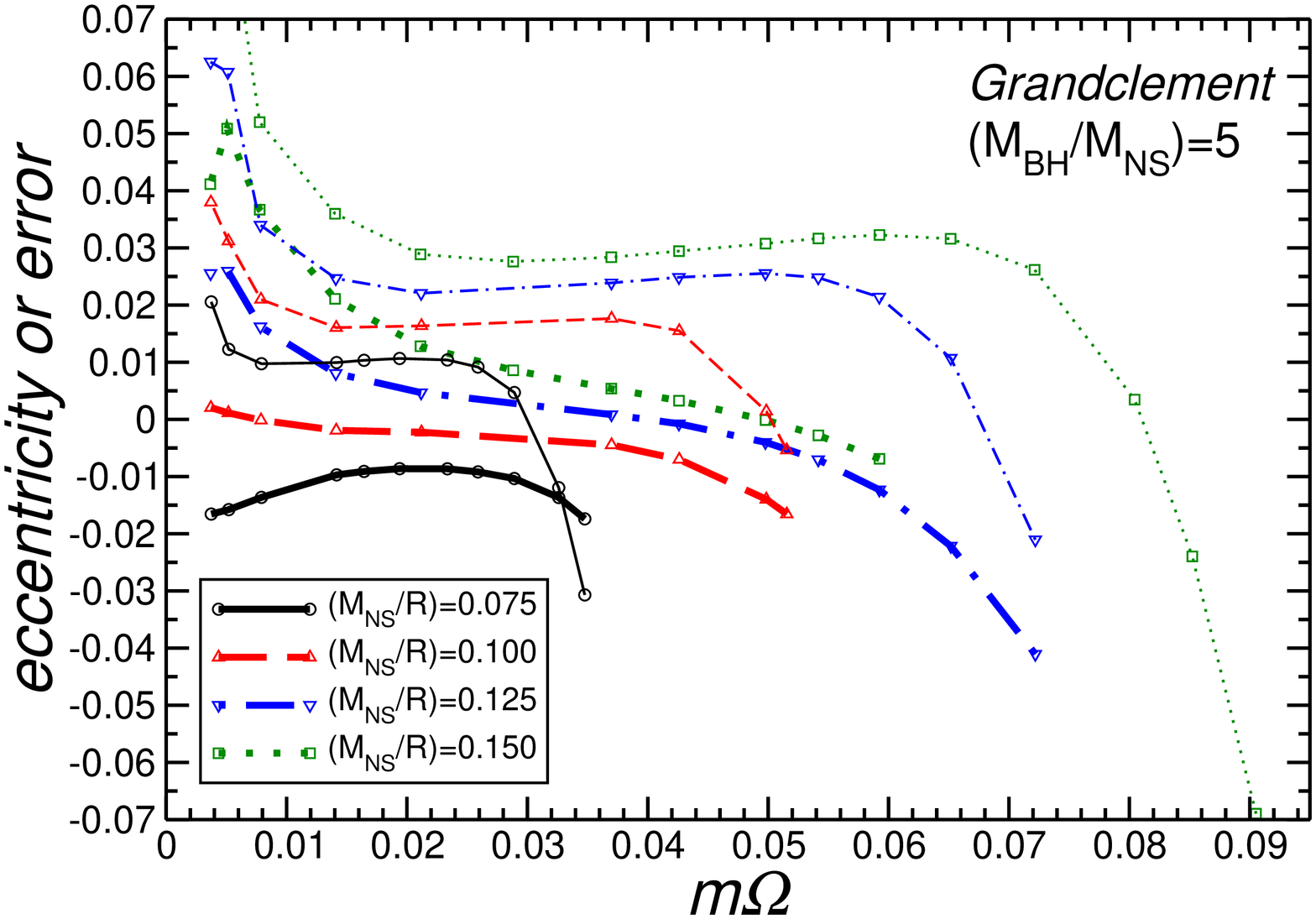,width=6cm,angle=-90} \\
\end{tabular}
\caption{
Best-fit eccentricity estimated from the energy (thick lines) and the
angular momentum (thin lines) for irrotational, black 
hole-neutron star binaries. Left: improved data
from Taniguchi {\em et al.} \cite{Taniguchi:PC}, for a compactness factor of
0.145, and four mass ratios.
Right: data from Grandcl\'ement, for a single mass ratio of 5:1, and a
range of compactness factors.  The dotted curves in each figure
correspond to a mass ratio 5 and similar compactness factors; they
also correspond to the data plotted in Fig. \ref{fig:EJ-nsbh}.
\label{fig:ecc-nsbh}
  }
\end{center}
\end{figure*}

Figure \ref{fig:EJ-nsbh} shows the gross agreement between the
Taniguchi {\em et al.} and Grandcl\'ement data, for the one case (mass
ratio 5, compactness $\sim 0.15$) where they can be compared, and
between the data and the corresponding PN diagnostics.  Figure
\ref{fig:ecc-nsbh} shows the inferred eccentricities (or the
accuracies of the PN comparisons) for all the NS-BH data sets.  The
eccentricities are somewhat larger than in the NS-NS case (and
significantly larger for the Grandcl\'ement mass ratio $5:1$ data), and
vary more strongly as a function of $m\Omega$, especially near the
end-points of the sequences, where the larger tidal deformations cause
numerical errors.   In addition, the eccentricities (or errors) 
appear to increase
with mass ratio.  This may not be surprising, as both numerical groups
acknowledge that numerical errors tend to increase with mass ratio.  
At the same time, the convergence of the PN series depends on
mass ratio, being best for equal masses, and rather poor for extreme
mass ratios (see Fig. 1 of Ref.~\cite{MW2}).
In the equal-mass case, where the numerical errors are smallest and
the PN series converges the best, the
inferred eccentricities are small, less than 0.01 from the
energy.  

Because these NS-BH numerical simulations are still in their infancy
compared to the BH-BH and NS-NS simulations, it is premature to draw
definite conclusions from comparisons with PN theory.  Instead, a PN
diagnostic may be a useful tool in testing improved simulations.  For
example, in comparing eccentricities inferred from the original data
published in \cite{Taniguchi:2007xm} with those inferred from the 
unpublished data \cite{Taniguchi:PC} that used an
improved black-hole zero spin condition (plotted in the left panel of
Fig. \ref{fig:ecc-nsbh})  we found a significant reduction in the
disagreement between values of $e$ inferred from $E$ and those
inferred from $J$, especially at the closer separations.  In Fig.
\ref{fig:ecc-nsbh} (left panel) the eccentricities from $E$ and $J$
closely track each other for the mass ratios 1, 2 and 3, both
increasing slightly with $m\Omega$; with the earlier data, the
eccentricities inferred from $J$ actually grew strongly with $m\Omega$,
diverging from those inferred from $E$.  A similar improvement
resulting from the new black-hole spin condition was noted in BH-BH
simulations (see \cite{Berti:2006bj}, Fig. 2).  

\section{Conclusions}
\label{sec:Conclusion}

We have used a post-Newtonian diagnostic tool to examine initial data
sets for non-spinning, double neutron star and neutron star-black hole
binary systems.  The PN equations included the effects of tidal
interactions, parametrized by the compactness of the neutron stars and
by suitable values of apsidal constants.  

We found that PN theory
agrees quite well with the NS-NS initial data, typically to better
than half a percent except where tidal distortions are becoming
extreme.  The differences between PN theory and the numerics are
sufficiently large and systematic that they could be interpreted as
representing residual eccentricity in the initial orbits.  It will be
interesting to see if simulations of the time-evolution from these
initial data sets show signs of eccentricity in the orbits.

Comparison of our PN diagnostic with NS-BH initial data sets showed
poorer agreement, not inconsistent with the larger numerical errors
present in these preliminary results.  We argued that a PN diagnostic
could be a useful tool for examining future improved data for NS-BH
binaries.

Tidal effects depend on the values of the apsidal constants, and we
addressed how relativistic neutron-star structure 
would affect their values.  An open question is whether our estimate
of a relativistic $k_2$ from rotational deformations 
is a reasonable estimate of
the tidally-induced $k_2$.  This would be an interesting and important
problem in relativistic stellar structure.

\acknowledgments 
We thank Greg Cook, Eric Gourgoulhon, Philippe Grandcl\'ement,
Mark Miller, Wai-Mo Suen and Keisuke
Taniguchi 
for useful discussions and for sharing with us their numerical data.
This work was partially funded by the National Science Foundation under grant
numbers PHY 03-53180 and PHY 06-52448.  CMW and EB are grateful to the
Institut d'Astrophysique de Paris, Universit\'e Pierre et Marie Curie 
for their hospitality
during the final stages of this work.

\appendix

\section{Tidal effects in relativistic neutron stars}
\label{sec:rotcorr}

In \cite{MW2}, we argued that, for compact bodies such as neutron
stars or black holes, the effects of finite size, such as rotational
kinetic energy, tidal distortions, and so on, would be of effectively
higher PN order than expected a priori, because the bodies' sizes scale
as their masses $M$, i.e. 
$R \sim q\,M$, where $q$ is a factor of order unity.
Thus, for example, the rotational kinetic
energy of a body can be expressed as 
$E_{\rm Rot}\sim I\omega^2/2 \sim MR^2 (M/r^3)(\omega/\Omega)^2
\le E_N q^2 (M/r)^2$, where $E_N$ is the Newtonian orbital energy, and
$\omega$ is the body's rotational angular velocity.
For compact bodies that are rotating no faster than the orbital
angular velocity, rotational kinetic energy
can thus be viewed as an effectively 2PN contribution
to the energy, even though formally it is a Newtonian effect. 

Similarly, tidal deformations have the dependence $E_{\rm Tidal} \sim
(M^2/R)(R/r)^6 \sim E_N q^5 (M/r)^5$, and so are effectively 5PN
order.  However, for neutron stars, $q$ can be as large as 8, and so
the effects could be important, as we have seen.  

While we have argued that the smallness of these effects justifies a
Newtonian treatment of tidal and rotational deformations~\cite{MW2}, 
one might
worry that, in relativistic 
neutron stars, the values of such quantities as apsidal constants could
differ strongly from their Newtonian values.  In this
appendix, we address this question.  

Unfortunately, it is difficult to treat tidal effects in relativistic
stars,
because that would involve constructing a relativistic stellar
model in the gravitational field of another body, which is exactly the
problem of inspiralling binaries treated by numerical relativity.

On the other hand, there is now a rather complete description of
isolated rotating relativistic stars.  In Newtonian gravity, there is
no distinction between rotational flattening and tidal distortion,  
at least for $l=2$ linear perturbations,
and thus a single apsidal constant $k_2$ applies to both
distortions (the only difference is that one distortion is oblate
while the other is prolate).  For higher $l$ this degeneracy breaks
down.  In relativity, by contrast, there is a difference in principle
between rotational and tidal
deformation for $l=2$, even for linear perturbations, because of the
presence in rotating stars 
of such phenomena as frame-dragging and the relativistic contributions of
rotational kinetic energy to self-gravity.  Nevertheless, for the
slowly rotating configurations that we expect to be relevant to
estimating apsidal constants for small distortions, we might
assume that the $k_2$ that governs rotationally induced distortions in
a relativistic star might not be too different from the $k_2$ that
governs tidal distortions.  This permits one to read off $k_2$ from
the quadrupole moment induced by the slow rotation. 

Because the effects of $k_3$ were extremely small in our diagnostic, 
we will not
consider this apsidal constant further.  Therefore, in this appendix
we will endeavor to estimate $k_2$ for relativistic neutron stars and
compare the results with those derived from Newtonian gravity.
First we review the calculation of Newtonian apsidal constants.

\subsection{Apsidal constants in Newtonian gravity}
\label{sec:apsidaln}

\begin{table}[htb]
\centering
\caption{Newtonian apsidal constants 
  (compare with Table II in \cite{BO} and Table II in \cite{K1}); $l$
is the angular harmonic index, $\Gamma$ and $n$ are adiabatic indices,
with $\Gamma =1+1/n$. }
\vskip 12pt
\begin{tabular}{@{}cccccccc@{}}
\hline
\hline
$l$ &$n=2/3$ &$n=0.80$ &$n=1.00$ &$n=1.25$ &$n=1.50$ &$n=2.00$ &$n=2.50$ \\
&$\Gamma=2.5$&$\Gamma=2.25$&$\Gamma=2$&$\Gamma=1.8$&$\Gamma=1.67$&$\Gamma=1.5$&$\Gamma=1.4$ \\
\hline
2 &0.375966 &0.325098 &0.259909 &0.194339 &0.143279  &0.073938 &0.034852  \\
3 &0.164696 &0.138660 &0.106454 &0.075590 &0.052849  &0.024394 &0.010192   \\
4 &0.098546 &0.081155 &0.060241 &0.040967 &0.027393  &0.011508 &0.004342  \\
5 &0.067424 &0.054485 &0.039293 &0.025748 &0.016569  &0.006420 &0.002220  \\
6 &0.049791 &0.039575 &0.027827 &0.017651 &0.010984  &0.003966 &0.001272  \\
7 &0.038649 &0.030270 &0.020810 &0.012824 &0.007745  &0.002628 &0.000789 \\
\hline
\hline
\end{tabular}
\label{tab:apsidaln}
\end{table}

In Newtonian gravity, apsidal constants can be computed by a standard
method for different
polytropic indices $n$ or $\Gamma=1+1/n$ 
and different values of $l$ using the numerical method
described in \cite{BO}. The results are
listed, for example, in the classical monographs by Kopal \cite{K1,K2}.
Unfortunately these works do not list apsidal constants for $n=2/3,~0.8$ and
$1.25$. Since these values are needed for neutron
stars, we report here results of our own numerical
calculations.  In Newtonian gravity, apsidal constants are defined by the
relation (see Appendix B of \cite{MW2} for a detailed discussion)
\be
k_l=\f{l+1-\eta_l(R)}{2l+2\eta_l(R)}\,,
\ee
where $R$ is the stellar radius and the function $\eta_l$ is a solution of the
Clairaut-Radau differential equation
\be\label{clairaut}
r \eta'_l+\eta_l(\eta_l-1)+6D(\eta_l+1)=l(l+1)\,,
\ee
with initial condition $\eta_l(0)=l-2$. A prime denotes a derivative with
respect to the distance $r$ from the center of the star; the quantity $D\equiv
\rho(r)/\bar \rho(r)$, where $\rho(r)$ is the density of the undistorted
configuration at $r$ and $\bar \rho(r)$ is the mean density inside the volume
of radius $r$. Of course, to obtain the density profile we must specify an
equation of state. The simplest choice is a (Newtonian) polytropic EOS
$\rho=\rho_c y^n$, where $\rho_c$ is the central density, $n$ is the
polytropic index and $y$ satisfies the Lane-Emden differential equation
\be\label{lane}
y''+\f{2}{r}y'+y^n=0\,,
\ee
subject to the initial conditions $y(0)=1$, $y'(0)=0$. We integrated the
system (\ref{clairaut}-\ref{lane}) following the method described in
\cite{BO}. Table \ref{tab:apsidaln} shows the results. When a comparison with
Table II of \cite{BO} is possible, the agreement is usually at the level of
five to six decimal places or better.

\subsection{Apsidal constant $k_2$ for relativistic stars}
\label{sec:apsidalHT}

We now turn to rotating relativistic stars, with the goal of using
their rotational distortions to estimate $k_2$.
We use the general relativistic, slow-rotation Hartle-Thorne framework
\cite{H,HT}, as
described in \cite{BWMB}. 
Rotating stars can be characterized by a dimensionless parameter
$\epsilon$, given by 
\be
\epsilon = \omega \sqrt{R^3/M} \,,
\label{epsilon}
\ee
which is the ratio between the body's angular velocity and the
Keplerian angular velocity at the equator of a non-rotating stellar model
(this $\epsilon$ should not be
confused with the post-Newtonian expansion parameter).
Even for the fastest-rotating configurations (that is, when we
consider corotating binaries at the ISCO)
$\epsilon\lesssim 0.35$ (see the last column of Table \ref{tab:etafit2}), so a
slow-rotation approximation is valid for our purposes.  For
constant-baryonic mass sequences, the Hartle-Thorne slow-rotation
approximation induces an error with respect to the ``true'' value of the
quadrupole moment which depends on the stellar mass and on the EOS. Since the
apsidal constant $k_2$ is proportional to the star's quadrupole moment $Q$
(see below), our calculation is valid as long as the error in $Q$ induced by
the slow-rotation approximation is small.  In Ref.~\cite{BWMB} we compared the
Hartle-Thorne approximation with rotating neutron star calculations in full
general relativity (see eg. Fig.~1 in \cite{BWMB}). 
From that comparison we conclude that our
Hartle-Thorne based values of $Q$ are good to within a few percent of
the ``true'' values,  which is more than acceptable for our estimates.

Following standard conventions (see eg. \cite{stergioulas}), 
we use the relativistic polytropic EOS
$p=\kappa\rho^\Gamma$, with energy density $\epsilon=\rho +{P}/(\Gamma-1)$,
and work in polytropic units.
For $n\simeq
0.5-1$ one obtains models with bulk properties comparable with those of
observed neutron stars; polytropic indices $n<1$ ($n>1$) yield stiff (soft)
stellar models, respectively. 
We are interested in
fitting numerical sequences of quasiequilibrium binary systems
containing neutron stars, which
are computed at constant baryonic mass.  Correspondingly, we consider
rotational corrections along sequences of (isolated) rotating
stars with constant baryonic mass.

Given a sequence of rotating models, we can
compute the ``relativistic'' apsidal constant $k_2$, defined as
\be
k_2=\f{3}{2}\f{Q}{R^5\omega^2}=\f{3}{2}\f{Q}{MR^2}\,.
\ee
for fixed values of $M/R$, where $Q$
is the quadrupole moment of the star as determined from the exterior 
geometry. 
The last step makes use of the fact that, in the
Hartle-Thorne formalism, $Q$ and all rotation-dependent quantities (scaling
with some power of $\omega$) are computed at the ``reference'' angular
velocity $\omega=\sqrt{M/R^3}$.  As shown in Appendix B.3 of \cite{MW2}, 
this is the most natural relativistic generalization of the Newtonian apsidal
constant. In fact, we verified that in the Newtonian limit $M/R\to 0$ the
Hartle-Thorne code reproduces the Newtonian apsidal constants listed in Table
\ref{tab:apsidaln}.  Notice that since $Q\propto \epsilon^2$, and $k_2\propto
Q/\omega^2$, $k_2$ does not depend on the rotation rate, but only on $M/R$.
A concise way of presenting the data is by a polynomial least-squares fit
(basically a post-Newtonian series) of the form
\be
k_2=\sum_{j=0}^2 k_2^{(j)}(M/R)^j \,.
\label{k2rel}
\ee
The coefficients $k_2^{(j)}$ are listed in Table \ref{tab:apsidalMRfit}. We
also report the percentage errors of the fits with respect to the numerical
calculations, $\Delta k_2^{\rm max}={\rm max}[(k_2-k_2^{\rm num})/k_2^{\rm
  num}]$: the fit is in excellent agreement with the numerical data for all
values of $n$ and in the whole range of $M/R$.  
For this reason, we used the analytical fits when computing the
relativistically corrected apsidal constants to use in the diagnostic
equations.

Over the range of
compactness factors shown, $k_2$ decreases by a factor of order two.
This gives a small but noticeable shift in the eccentricities inferred
from the numerical data, as displayed in Figs. \ref{201212-D201212}
and \ref{201818-D201818}; because the diagnostic baseline shifts
toward a point-mass model (smaller $k_2$) the data referred to that
baseline (the triangles) shift upward.

\begin{table}[htb]
\centering
\caption{Fitting coefficients for the ``relativistic'' apsidal
  constants as defined in Eq. (\ref{k2rel}).
The fifth column shows
the maximum percentage error of the fit with
  respect to the numerical data, and the last column shows the
maximum value of $M/R$ used in the fit.
}
\vskip 12pt
\begin{tabular}{@{}cccccc@{}}
\hline
\hline
$n$     &$k_2^{(0)}$ &$k_2^{(1)}$ &$k_2^{(2)}$ &$\Delta k_2^{\rm max} (\%)$
&$(M/R)^{\rm max}$\\
\hline
1.25    &0.194339   &-0.773249  &0.733840   &0.32  &0.172\\
1.00    &0.259909   &-0.956592  &1.052813   &0.21  &0.214\\
0.80    &0.325098   &-1.144687  &1.377584   &0.23  &0.252\\
2/3     &0.375966   &-1.296016  &1.634947   &-0.80 &0.278\\
\hline
\hline
\end{tabular}
\label{tab:apsidalMRfit}
\end{table}

\subsection{Mass and moment of inertia of relativistic stars}
\label{massmoment}

For future applications of the PN diagnostic to binaries with rotating
neutron stars, it will be useful to have estimates of the variation in
the mass and angular momentum of isolated rotating neutron stars, as a
function of compactness and angular velocity.

We used the code described in~\cite{BWMB} to compute rotational corrections to
the gravitational mass of the isolated neutron star
as a function of rotation for different values of $n$. We assume
that the non-rotating model has one of the compactness $M/R$ 
values chosen in Table \ref{tab:emmodels};
this fixes the value of the baryonic mass for our nonrotating model. Then we
increase the rotation rate at fixed baryonic mass (along an ``evolutionary
sequence'') and compute the rotationally corrected gravitational mass 
\be
M_{\rm rot} \equiv (1+\beta)M_{\rm nonrot} \,.
\label{Meta}
\ee
More details on our numerical procedure are given in
Sec.~3.1 of \cite{BWMB} (the main difference being that there we fix the
angular momentum and gravitational mass, here we fix angular velocity and
baryonic mass).

For each value of $n$ and of the nonrotating star's compactness $(M/R)$, 
we compute 11 models
(the nonrotating model and 10 models equispaced in the angular velocity
$\omega$).  Then we perform a polynomial least-squares fit of $\beta$ as a
function of the star's angular velocity $\bar{\omega}$ (expressed in polytropic
units):
\be
\beta=\sum_{k=1}^3 \beta^{(k)}\bar{\omega}^k\,.
\label{etasum}
\ee
The fitting coefficients $\beta^{(k)}$ are listed in Table \ref{tab:etafit2}.

\begin{table}[htb]
\centering
\caption{Fitting coefficients for $\beta(\omega)$ as defined in Eqs.
(\ref{Meta}) and (\ref{etasum}). $(M/R)_{\epsilon=0}$ is the
  compactness and $M$ is the gravitational mass 
(in polytropic units) of the nonrotating model. 
The seventh column shows
the maximum percentage error of the fit with
  respect to the numerical data, and the last two columns show the
maximum value of $\bar{\omega}$ or $\epsilon$ used in the fit.
  }
\vskip 12pt
\begin{tabular}{@{}ccccccccc@{}}
\hline
\hline
$n$  &$(M/R)_{\epsilon=0}$ &$M$  &$10^3\times \beta^{(1)}$ &$\beta^{(2)}$
&$\beta^{(3)}$ &$\Delta \beta^{\rm max} (\%)$ 
&$\bar{\omega}^{\rm max}$ &$\epsilon^{\rm max}$\\
\hline
1.25 &0.08 &0.143165 &1.500742 &0.321866 &1.617628 & 0.98 &0.048 &0.316\\
1.25 &0.10 &0.164915 &0.138509 &0.325523 &0.740812 & 0.49 &0.058 &0.316\\
1.25 &0.12 &0.182153 &0.277598 &0.272061 &0.736909 & 0.46 &0.071 &0.327\\
1.25 &0.14 &0.194658 &1.062211 &0.207540 &0.858382 & 0.78 &0.083 &0.329\\
\hline
1.00 &0.12 &0.122286 &-0.072168&0.153656 &0.082648 & 2.61 &0.109 &0.331\\
1.00 &0.14 &0.136233 &0.628680 &0.126834 &0.168151 & 0.31 &0.125 &0.339\\
1.00 &0.16 &0.147838 &0.781091 &0.113525 &0.172454 & 1.69 &0.143 &0.347\\
1.00 &0.18 &0.156736 &0.518670 &0.105391 &0.156755 & 1.32 &0.160 &0.350\\
\hline
0.80 &0.12 &0.087240 &-0.221682&0.087986 &0.013138 & 0.78 &0.152 &0.327\\
0.80 &0.14 &0.099987 &0.534238 &0.076339 &0.051296 &-0.57 &0.171 &0.335\\
0.80 &0.16 &0.111652 &0.278734 &0.078449 &0.034777 &-0.53 &0.191 &0.343\\
0.80 &0.18 &0.122021 &0.345399 &0.075647 &0.035287 &-0.24 &0.208 &0.345\\
\hline
2/3  &0.14 &0.080637 &0.794037 &0.049246 &0.034088 & 0.59 &0.211 &0.332\\
2/3  &0.16 &0.091591 &-0.248756&0.061873 &-0.000499&-0.37 &0.232 &0.340\\
2/3  &0.18 &0.101844 &0.009904 &0.059124 &0.008106 &-0.43 &0.253 &0.347\\
2/3  &0.20 &0.111209 &0.370593 &0.055858 &0.014234 &-0.15 &0.274 &0.351\\
\hline
\hline
\end{tabular}
\label{tab:etafit2}
\end{table}

Relativistic corrections to the Newtonian moment of inertia,
may be useful for the diagnosis of binaries containing rotating
bodies, in order to
evaluate the total angular momentum (including that of
the stars themselves) and such effects as spin-orbit coupling.
Following \cite{MW2}
we define a dimensionless factor
\be
\alpha\equiv I/(MR^2)=(R_g/R)^2\,.
\label{Ialpha}
\ee
The ``radius of gyration'' is defined as $R_g=(I/M)^{1/2}$. In the
Hartle-Thorne formalism, to leading order, $R_g$ does not depend on the
rotation rate (see the discussion following Eq.~(11) of \cite{BWMB}). 
In Table
\ref{tab:alphafit} we provide fitting coefficients for a post-Newtonian
inspired power-series expansion for $\alpha$, as computed using the
Hartle-Thorne formalism:
\be
\alpha=\sum_{j=0}^2 \alpha^{(j)}(M/R)^j\,.
\label{alphasum}
\ee

\begin{table}[htb]
\centering
\caption{
Fitting coefficients for $\alpha(M/R)$ as defined in Eqs.
(\ref{Ialpha}) and (\ref{alphasum}). 
The fifth column shows
the maximum percentage error of the fit with
  respect to the numerical data, and the last column shows the
maximum value of $M/R$ used in the fit.
}
\vskip 12pt
\begin{tabular}{@{}cccccc@{}}
\hline
\hline
$n$     &$\alpha^{(0)}$ &$\alpha^{(1)}$ &$\alpha^{(2)}$ 
&$\Delta \alpha^{\rm max} (\%)$ &$(M/R)^{\rm max}$\\
\hline
1.25    &0.231893 &0.256949 &0.065809 &0.19  &0.172\\
1.00    &0.261297 &0.274488 &0.349978 &0.15  &0.214\\
0.80    &0.286378 &0.275807 &0.569004 &-0.07 &0.252\\
2/3     &0.303974 &0.269094 &0.714660 &-0.19 &0.278\\
\hline
\hline
\end{tabular}
\label{tab:alphafit}
\end{table}

\end{document}